\documentclass[preprint2]{aastex701}

\defcitealias{2023_Vesa}{Paper\,I}
\newcommand{\IBISCa}[1]{Ca\,{\sc ii}\,8542}
\newcommand{\HMI}[1]{Fe\,{\sc i}\,6173}
\newcommand{\IBISFeLowerPhot}[1]{Fe\,{\sc i}\,7090}
\newcommand{\IBISFeUpperPhot}[1]{Fe\,{\sc i}\,5434}
\newcommand{\IBISK}[1]{K\,{\sc i}\,7699}
\newcommand{\DeubnerCi}[1]{C\,{\sc i}\,5380}
\newcommand{\DeubnerTi}[1]{Ti\,{\sc ii}\,5381}
\newcommand{\DeubnerNa}[1]{Na\,{\sc i\,D$_{1}$}\,5896}
\usepackage{caption}
\usepackage{float}
\usepackage{subfigure}
\usepackage{url}
\usepackage{amsmath}
\DeclareCaptionFormat{cont}{#1 (cont.)#2#3\par}

\usepackage{gensymb}

\graphicspath{{./}{Figures/}}

\begin{document}

\title{Atmospheric Gravity Waves Modulated by the Magnetic Field Configuration}

\correspondingauthor{Oana Vesa} 

\author[orcid=0000-0001-6754-1520,gname=Oana,sname=Vesa]{Oana Vesa}
\affiliation{W. W. Hansen Experimental Physics Laboratory, Stanford University, Stanford, CA 94305-4085, USA}
\email[show]{ovesa@stanford.edu}  

\author[orcid=0000-0001-7525-7423,gname=Julio,sname=Morales]{Julio Morales}
\affiliation{Department of Astronomy, New Mexico State University, P.O. Box 30001, MSC 4500, Las Cruces, NM 88003-8001, USA}
\email{jmmorale@nmsu.edu}

\author[orcid=0000-0001-9659-7486, gname=Jason,sname=Jackiewicz]{Jason Jackiewicz}
\affiliation{Department of Astronomy, New Mexico State University, P.O. Box 30001, MSC 4500, Las Cruces, NM 88003-8001, USA}
\email{jasonj@nmsu.edu}

\author[orcid=0000-0002-9820-9114, gname=Gangadharan,sname=Vigeesh]{Gangadharan Vigeesh}
\affiliation{Institut für Sonnenphysik (KIS), Georges-Köhler-Allee 401a, D-79110 Freiburg, Germany}
\email{vigeesh@leibniz-kis.de}

\author[orcid=0000-0001-8016-0001, gname=Kevin,sname=Reardon]{Kevin Reardon}
\affiliation{National Solar Observatory, Boulder, CO 80303, USA}
\email{kreardon@nso.edu}

\shortauthors{Vesa et al.}
\shorttitle{AGWs and Magnetic Fields}
 
\begin{abstract}  % 250 words max
Atmospheric gravity waves (AGWs) are buoyancy-driven waves excited by turbulent convection and contribute to the dynamics and energy transport of the lower solar atmosphere. We present high-resolution, multi-wavelength observations from the Interferometric Bidimensional Spectrometer and the Solar Dynamics Observatory to investigate AGW behavior across different viewing geometries and magnetic field \added{configurations}. Using Fourier spectral analysis to compute phase differences and coherence spectra, we detect the signature of propagating AGWs carrying energy upwards at temporal and spatial scales consistent with theory, simulations, and prior observations. Although AGW behavior is modulated by the magnetic field \added{configuration}, particularly the field inclination, these effects are not highly discernible in our observed $k_{\rm{h}}-\nu$ phase difference diagrams. After filtering to isolate the AGW regime, we compute spatial coherence-weighted phase difference maps and examine binned coherence-weighted phase differences as functions of the field strength and inclination. Our results show that AGWs are efficiently suppressed and/or reflected in intermediate to strong, vertically oriented fields in the upper photosphere, while they propagate rather freely in QS and transverse fields. These findings agree with a simulated vertical 100\,G field using CO$^{5}$BOLD. Simulated $k_{\rm{h}}-\nu$ phase differences derived from a 3D magnetohydrodynamic dispersion relation also qualitatively agree with our upper photospheric IBIS diagnostics and reinforce that the magnetic field \added{configuration} modulates the propagation of AGWs. This work demonstrates the potential of AGWs as magneto-seismology diagnostics for probing average magnetic field properties in the lower solar atmosphere.
\end{abstract}

\section{Introduction} \label{sec:intro}

The solar atmosphere is replete with oscillations of various spatio-temporal scales, including atmospheric gravity waves (AGWs).
AGWs are low-frequency (1--4\,mHz), buoyancy-driven waves copiously generated by turbulent convection in the stably stratified atmosphere \citep{1967_Lighthill, 1967_Stein}.
Propagating at an oblique angle throughout the lower solar atmosphere, they are expected to reach low to mid-chromospheric heights \citep{1981_Mihalas_Toomre, 1982_Mihalas_Toomre} and impart energy comparable to the chromospheric radiative losses \citep{2008_Straus_Fleck_Jeffries_etal, 2009_Straus_Fleck_Jeffries_Severino_Steffen_Tarbell}.

Observational evidence of AGWs dates back to the 1960s and 1970s, with analysis of velocity and intensity fluctuations suggesting their presence at low frequencies and moderate horizontal wavenumbers \citep{1968_Frazier2, 1974_Deubner}.
Subsequent phase analysis of ground- and space-based observations taken primarily near quiet-Sun (QS) disk center have unambiguously shown the signature of propagating AGWs carrying energy upward at the expected temporal and spatial scales \citep[e.g.,][]{1989_Deubner_Fleck, 2023_Kumar_Jumar_Rajaguru_Mathew_Bayanna}, corroborating numerical simulations and synthesized observations \citep[e.g.,][]{2017_Vigeesh, 2020_Vigeesh_Roth}, numerical models \citep[e.g.,][]{2009_Newington_Cally, 2016_Hague_Erdelyi}, and linear theory \citep[e.g.,][]{1981_Mihalas_Toomre, 1982_Mihalas_Toomre}.
In all of these observations, a key characteristic of propagating AGWs carrying energy upward is noted: phase and group velocities are nearly orthogonal, with the vertical component of the energy (group velocity) propagating in the opposite direction to the vertical phase velocity \citep{1960_Hines}.

As magnetic fields of various spatio-temporal scales pervade the solar atmosphere, interactions with AGWs are highly probable and have garnered considerable attention over the years \citep[e.g.,][]{2011_Newington_Cally, 2020_Vigeesh_Roth, 2023_Kumar_Jumar_Rajaguru_Mathew_Bayanna}.
The earliest discussion on this topic can be traced to \citet{1967_Lighthill}, who theorized that AGWs can efficiently transform into Alfv\'{e}n waves at higher atmospheric heights in an inclined magnetic field \added{geometry}.
Theoretical work by \citet{1974_Stein_Leibacher} also suggested that magnetic fields could suppress the propagation of AGWs with large horizontal wavenumbers.

In-depth investigations into the influence of magnetic fields on AGWs began with observational work by \citet{2008_Straus_Fleck_Jeffries_etal} who suggested suppression and possible mode conversion of AGWs into Alfv\'{e}n waves at locations of magnetic flux in a QS disk center region.
Subsequent numerical models confirmed that even weak vertical ($B = 10$\,G) or inclined magnetic fields can significantly alter their behavior, reflecting them as slow magnetoacoustic waves well below the plasma-$\beta=1$ level or facilitating mode conversion depending on the field geometry \citep{2009_Newington_Cally, 2011_Newington_Cally, 2016_Hague_Erdelyi}.
Realistic numerical simulations of the solar atmosphere further demonstrated that strong vertical fields suppress AGW propagation in the upper photosphere while horizontal fields permit their transmission to chromospheric heights \citep{2017_Vigeesh, 2019_Vigeesh_Roth_Steiner_Jackiewicz, 2021_Vigeesh_Roth_Steiner_Fleck}.
However, recent synthetic observations suggest that the effects of magnetic fields on AGWs remain challenging to detect observationally \citep{2020_Vigeesh_Roth}.

Overall, the properties of AGWs remain empirically under-constrained due to the limited number of observations across varying viewing geometries and magnetic field configurations. 
Their strong horizontal velocity components make limb observations particularly advantageous \citep{1981_Mihalas_Toomre}.
Numerical simulations also affirm their diagnostic potential for probing average QS magnetic fields in the upper photosphere, an atmospheric region accessible with high-resolution, high-cadence narrowband observations by the Interferometric BIdimensional Spectrometer\,\citep[IBIS;][]{Cavallini_IBIS}.

\begin{figure*}[ht!]
    \centering
    \includegraphics[width=\linewidth]{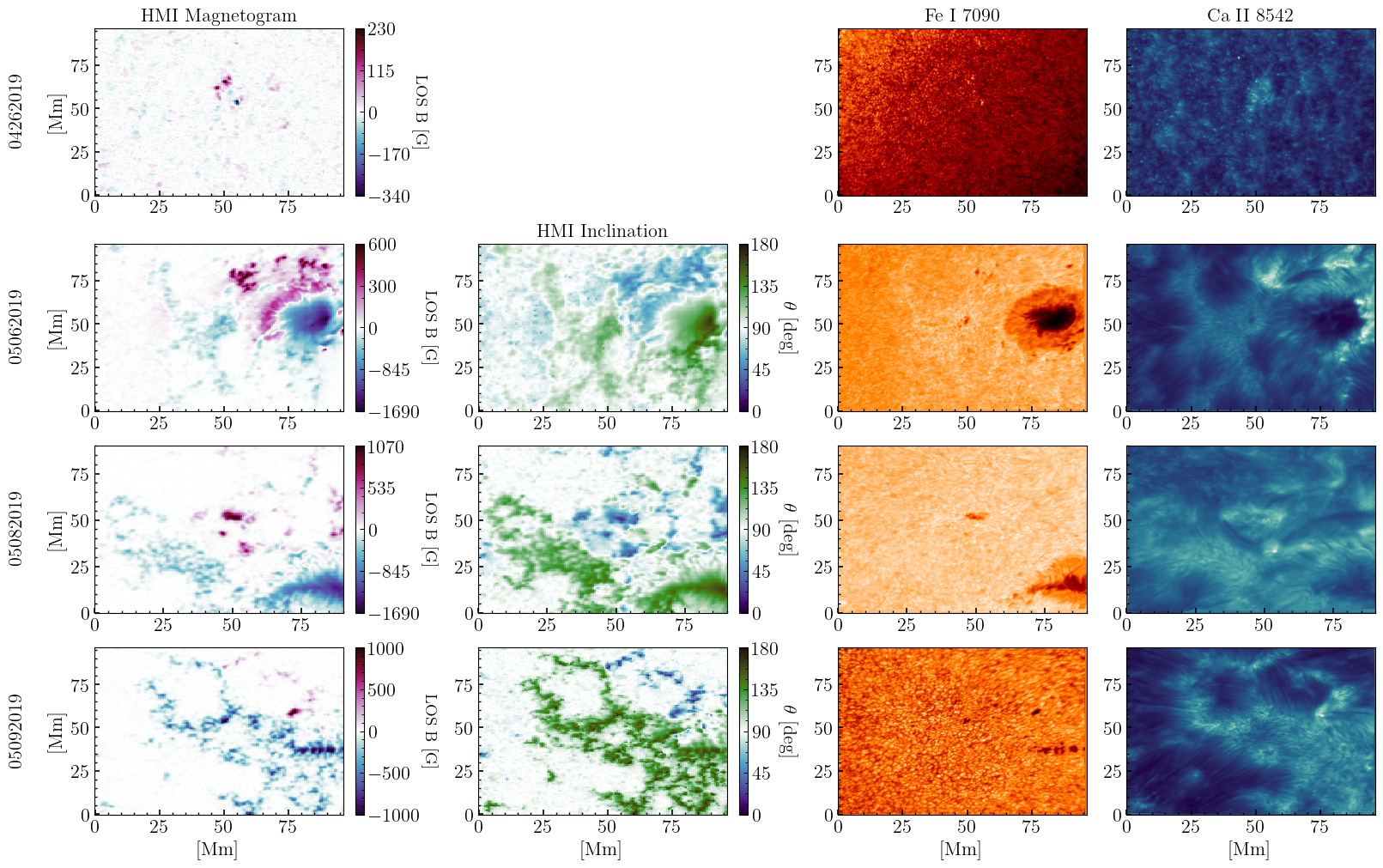}
    \caption{Co-spatial and co-temporal snapshots of the observations taken on 26 April 2019 (DS1), 06 May 2019 (DS2), 08 May 2019 (DS3), and 09 May 2019 (DS4). First column: HMI line-of-sight magnetogram showing the lower photospheric magnetic field. Second column: HMI SHARP magnetic field inclination, where transverse fields are shown in white. Third column: Intensity blue wing of \IBISFeLowerPhot{} sampling the low photosphere. Fourth column: Intensity line core of \IBISCa{} sampling the middle chromosphere.}
    \label{fig:AGW2_reference_image}
\end{figure*}

In \citet[][hereafter \citetalias{2023_Vesa}]{2023_Vesa}, we explored the behavior of AGWs throughout the lower solar atmosphere in a QS region observed near disk center with $\mu$ = 0.99 (where $\mu=\cos\theta$ and $\theta$ is the angle from the central meridian).
Propagating AGWs carrying energy upward and phase downward were detected at the expected spatial and temporal scales using simultaneously observed line core Doppler velocity and line minimum intensity fluctuations.
To facilitate direct comparisons between the properties of AGWs seen at QS disk center, toward the QS limb, and in varying magnetic field \added{configurations}, we follow the methodology in \citetalias{2023_Vesa}.

This work is structured as follows.
Datasets and simulation properties are described in Sect.\,\ref{sec:data}.
Results and analysis are in Sect.\,\ref{sec:results}.
The discussion and conclusions follow in Sects.\,\ref{sec:discussion}--\,\ref{sec:conclusion}.

\section{Dataset Properties} \label{sec:data}

\subsection{Observations} \label{subsec:observations}

We present four high-resolution, multi-wavelength observations acquired in April and May 2019, taken at different viewing angles and magnetic field environments.
The datasets are as follows: 1) QS region near the west limb ($\mu$ = 0.57) for 3\,hr starting at 15:24:06\,UT on 26 April 2019; 2) AR near the east limb ($\mu$ = 0.64) for 3.7\,hr starting at 14:34:40\,UT on 06 May 2019; 3) AR off disk center ($\mu$ = 0.88) for 3.6\,hr starting at 13:58:30\,UT on 08 May 2019; and 4) AR slightly off disk center ($\mu$ = 0.95) for 3.7\,hr starting at 15:37:08\,UT on 09 May 2019.
The datasets have a cadence of $\approx$\,16\,s and will be referred to herein as DS1, DS2, DS3, and DS4, respectively.
Co-spatial and co-temporal contextual snapshots are presented in Fig.\,\ref{fig:AGW2_reference_image}.

\begin{deluxetable*}{ccccccccc}
\tabletypesize{\scriptsize}
\tablewidth{0pt}
\tablecaption{Dataset and spectral line properties. \label{tab:dataset_and_spectral_line_properties}}
\tablehead{
\colhead{} & \colhead{} &  \colhead{} & \multicolumn{4}{c}{IBIS} & \phantom{} & \multicolumn{1}{c}{HMI/SDO}\\
\cline{4-7}
\cline{9-9}
\colhead{Date} & \colhead{$\mu$} & \colhead{Parameters} & \colhead{Fe\,{\sc i}\,7090} & \colhead{K\,{\sc i}\,7699} & \colhead{Fe\,{\sc i}\,5434}  & \colhead{Ca\,{\sc ii}\,8542} &  \colhead{} & \colhead{Fe\,{\sc i}\,6173}
}
\startdata
% \phm{04262019} & {Cadence [s]} & 16.4 & 16.4 & 16.4 & 16.4  & \phn & 12.0 & 12.0 & 12.0 \\
{(DS1) 04262019} & 0.57 &  \phm{Time Delay [s]} & 3.8 & 7.4 & 11.9 & 0.0 & \phn  & \nodata \\
% \hline
% \phm{05062019} & {Cadence [s]} & 16.38 & 16.38 & 16.38 & 16.38  & \phn & 12.0 & 12.0 & 12.0 \\
{(DS2) 05062019} & 0.64 &  {Time Delay [s]} & 4.4 & 7.4 & 11.6 & 0.0 & \phn  & \nodata \\
% \hline
% \phm{05082019} & {Cadence [s]} & 16.37 & 16.37 & 16.37 & 16.37  & \phn & 12.0 & 12.0 & 12.0 \\
{(DS3) 05082019} & 0.88 &\phm{Time Delay [s]} & 5.1 & 8.4 & 11.5 & 0.0 & \phn  & \nodata\\
% \hline
% \phm{05092019} & {Cadence [s]} & 16.38 & 16.38 & 16.38 & 16.38  & \phn & 12.0 & 12.0 & 12.0 \\
{(DS4) 05092019} & 0.95 & \phm{Time Delay [s]} & 3.8 & 7.4 & 11.3 & 0.0 & \phn  & \nodata \\
\hline
\phm{05092019} & \phm{05092019}  & {${\rm g_{\rm eff} }$} & 0.0 & 1.3 & 0.0 & 1.1 & \phn & 2.5  \\
% \phm{05092019} &  \phm{05092019} & {Formation Height [km]} & 200--250$^{1,2}$ & 450--650$^{3,4,5}$ & 500--650$^{6,7}$ & 1200--1500$^{2,3,8}$ &  \phn & 100--150$^{9,10}$ \\
\phm{05092019} &  \phm{05092019} & {Formation Height [km]} & 200--250\tablenotemark{a,b} & 450--650\tablenotemark{c,d,e} & 500--650\tablenotemark{f,g} & 1200--1500\tablenotemark{b,c,h} &  \phn & 100--150\tablenotemark{i,j}
\enddata
% \tablenotetext{a}{\cite{2008_Straus_Fleck_Jeffries_etal}} \tablenotetext{b}{\cite{2006_Janssen_Cauzzi}}
% \tablenotetext{c}{\cite{2017_QuinteroNoda}} % \tablenotetext{d}{\cite{1986_Severino_Roberti_Marmolino_Gomez}}
% \tablenotetext{e}{\cite{2007_Haberreiter}} % \tablenotetext{f}{\cite{2011_kneer_Bello}}
% \tablenotetext{g}{\cite{2010_BelloGonzalez}} % \tablenotetext{h}{\cite{2009_Reardon_Uitenbroek_Cauzzi}}
% \tablenotetext{i}{\cite{2011_Fleck}} % \tablenotetext{k}{\cite{2014_Nagashima}}
\tablerefs{(a) \cite{2008_Straus_Fleck_Jeffries_etal}; (b) \cite{2006_Janssen_Cauzzi}; (c) \cite{2017_QuinteroNoda}; (d) \cite{1986_Severino_Roberti_Marmolino_Gomez}; (e) \cite{2007_Haberreiter}; (f) \cite{2011_kneer_Bello}; (g) \cite{2010_BelloGonzalez}; (h) \cite{2009_Reardon_Uitenbroek_Cauzzi}; (i) \cite{2011_Fleck}; (j) \cite{2014_Nagashima}}
\tablecomments{Formation heights represent the average formation of the line core velocity signal above the photosphere at QS disk center. They are subject to change depending on viewing geometry and magnetic field environment.}
\end{deluxetable*}

These observations were acquired with IBIS, which was installed at the Dunn Solar Telescope (DST)\footnote[1]{\href{https://sunspot.solar/about/}{https://sunspot.solar/about/}} in Sunspot, New Mexico, in conjunction with the DST's high-order adaptive optics system \citep{2004_Rimmele_Richards_AO}. 
The absorption line profiles of \IBISFeLowerPhot{}, \IBISK{}, \IBISFeUpperPhot{}, and \IBISCa{} were sampled with a spatial sampling of 0{\farcs}096\,pixel$^{-1}$.
Table\,\ref{tab:dataset_and_spectral_line_properties} shows properties of the observed spectral lines, including the cadence and time delay between sequential sampling of the absorption line cores. 

These datasets are complemented by co-aligned space-based data from the Helioseismic and Magnetic Imager\,\citep[HMI;][]{HMI_instrumental} on board the Solar Dynamics Observatory\,\citep[SDO;][]{SDO_2012}.
HMI line-of-sight magnetograms and Dopplergrams were obtained using Rob Rutten's SDO alignment IDL package\footnote[2]{\href{https://robrutten.nl/Recipes_IDL.html}{https://robrutten.nl/Recipes\_IDL.html}}.
The data have a spatial sampling of 0{\farcs}6\,pixel$^{-1}$ and cadence of 12.0\,s.
In addition, DS2, DS3, and DS4 had corresponding Space-weather HMI Active Region Patches \citep[SHARPs;][]{2014_Hoeksema, 2014_Bobra}, which provided 720\,s magnetogram, field inclination, and field azimuth data products with a spatial sampling of 0{\farcs}5\,pixel$^{-1}$.

The data reduction for the IBIS observables, calculation of the line core Doppler velocities, accounting for the time delay between sampling, and rebinning process for both IBIS and HMI data to match (0{\farcs}6\,pixel$^{-1}$) were carried out following the methods described in \citetalias{2023_Vesa}.
Before the spatial interpolation, IBIS datasets had a Nyquist frequency of 30.5\,mHz, Nyquist wavenumber of 45.1\,Mm$^{-1}$, and wavenumber resolution of 0.09\,Mm$^{-1}$.
After the spatial interpolation, the datasets had a Nyquist wavenumber of 7.2\,Mm$^{-1}$ and wavenumber resolution of 0.09\,Mm$^{-1}$.
The frequency resolution is as follows: 90.2\,$\mu$Hz for DS1, 74.4\,$\mu$Hz for DS2, 76.3\,$\mu$Hz for DS3, and 74.4\,$\mu$Hz for DS4.

\subsubsection{Data Properties} \label{subsubsec:data_properties}
\begin{deluxetable}{ccccc}
% \tablenum{1}
\tabletypesize{\scriptsize}
\tablewidth{0pt}
\tablecaption{Magnetic field strength and inclination information from HMI. \label{tab:magnetic_field_information}}
\tablehead{
\colhead{Dataset} & \colhead{Mean} & \colhead{Median} & \colhead{RMS} & \colhead{(Min, Max)}}
\startdata
DS0 & 8.5 G & 4.2 G & 24.6 G & (-738 G, 687 G) \\
{} & \nodata & \nodata & \nodata & \nodata \\
\hline
DS1 & 7.3 G & 5.3 G & 12.2 G & (-391 G, 287 G) \\
{} & \nodata & \nodata & \nodata & \nodata \\
\hline
DS2 & 102.7 G & 19.8 G & 238.7 G & (-1862 G, 880 G) \\
{} & 92.7\degree & 91.2\degree & 94.3\degree & (24\degree, 169\degree) \\
\hline
DS3 & 90.0 G & 12.7 G & 219.3 G & (-1694 G, 1165 G)\\
{} & 91.3\degree & 91.4\degree & 93.1\degree & (14\degree, 152\degree) \\
\hline
DS4 & 51.4 G & 9.4 G & 120.5 G & (-1182 G, 1042 G) \\
{} & 101.1\degree & 93.4\degree & 103.0\degree & (4\degree, 179\degree)
\enddata
\tablecomments{DS0 refers to the QS disk center ($\mu$ = 0.99) dataset taken on 25 April 2019 described in \citetalias{2023_Vesa}.}
\end{deluxetable}

Contextual information on the magnetic field for each dataset can be found in Table\,\ref{tab:magnetic_field_information} and Fig.\,\ref{fig:AGW2_reference_image}.
DS1, taken near the west limb, is focused on a small pore ($\approx$1{\arcsec}), or ``micropore'' \citep{2019_Solovev_Parfinenko_Efremov_et_al}, surrounded by a bright rim that persists throughout the observations.
While the micropore has a median magnetic field value of 18\,G, we measure an unsigned line-of-sight magnetic field ($|B_{LOS}|$) RMS value of 12\,G and a median value of 5\,G for the entire field of view.
Therefore, it is comparable to the QS region in \citetalias{2023_Vesa} (DS0).
DS2, DS3, and DS4 are focused on active region (AR) 12740 during different evolutionary stages (and thus magnetic environments).
DS2 has the strongest average magnetic field with a $|B_{LOS}|$ RMS value of 239\,G, while DS4 has a $|B_{LOS}|$ RMS value of 120\,G.
Both sunspots and plage regions are largely associated with strong, vertically inclined fields \citep{2011_Borrero_Ichimoto, 2020_Pietrow_Kiselman_etal}, as seen in Fig.\,\ref{fig:AGW2_reference_image}.

\begin{figure*}[hbt!]
\centering
\includegraphics[width=\linewidth]{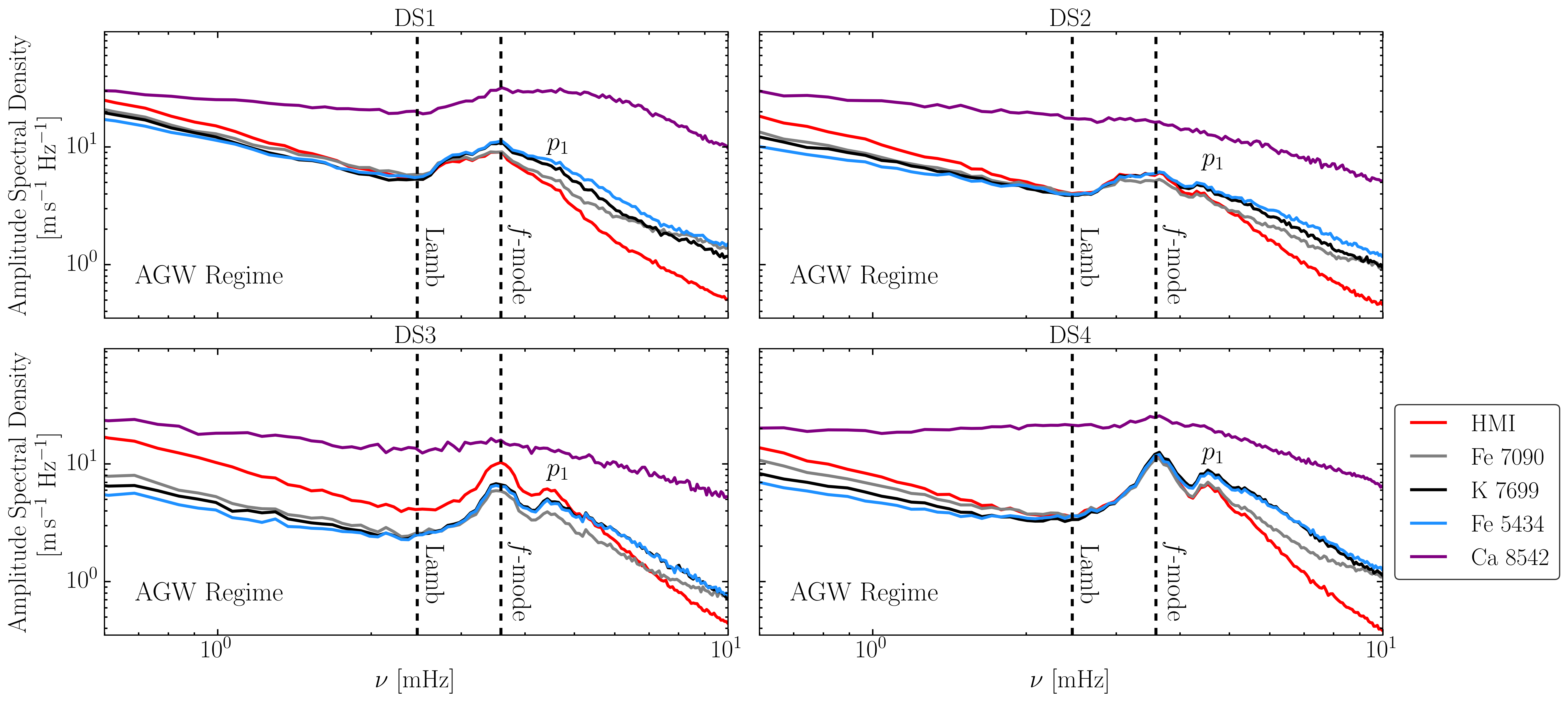}
\caption{Velocity amplitude spectral density profiles for IBIS and HMI (\HMI{}) at a horizontal wavenumber cut of 2.0\,Mm$^{-1}$. The AGW regime at low frequencies, the Lamb line ($\omega = c_sk_h$), the $f$-mode ($\omega^2 = gk_h$), and the first $p$-mode are labeled.} \label{fig:AGW2_amplitude_spectral_densities}
\end{figure*}

Figure\,\ref{fig:AGW2_amplitude_spectral_densities} shows the velocity amplitude spectral density profiles at a horizontal wavenumber cut of 2\,Mm$^{-1}$, illustrating the comparability of these data products.
The spectral density profiles of the photospheric Doppler diagnostics are qualitatively similar in all datasets, while the \IBISCa{} profiles (purple) are broader with larger amplitudes.
% The peaks associated with the $f$-mode and $p$-modes are not visible in the \IBISCa{} profiles.
\added{The peaks associated with the $f$-mode and $p$-modes are strongly visible in the photospheric lines, while they are much weaker and in most cases not discernible in the \IBISCa{} profiles.}

Comparable with \citetalias{2023_Vesa}, we note a qualitative match between the lower photospheric diagnostics, HMI (red) and \IBISFeLowerPhot{} (gray), and the upper photospheric diagnostics, \IBISK{} (black) and \IBISFeUpperPhot{} (blue), in the evanescent regime around the $p$-modes.
Towards the limb (DS1 and DS2), the $f$-mode and first $p$-mode peak become indistinguishable in the amplitude spectral density profiles.
This is consistent with work by \citet{1992_Deubner} showing that foreshortening affects $p$-modes and work by \citet{2013_Zhao_Chou} showing that $p$-modes within ARs are suppressed.

We observe strong velocity signals within the AGW regime for all datasets, which are comparable to the results in \citetalias{2023_Vesa}, demonstrating the presence of AGWs.
We note that at $\approx$\,1\,mHz, the velocity amplitudes are the largest for the QS limb (DS1) in line with expected theory proposed by \citet{1981_Mihalas_Toomre}.

\subsection{CO$^{5}$BOLD Simulation} \label{subsec:cobold_simulation}

Alongside observations, we utilize a high-resolution, realistic numerical simulation of the lower solar atmosphere constructed with CO$^{5}$BOLD \citep{2012_Freytag_COBOLD}, which was described in detail in \citet{2019_Vigeesh_Roth_Steiner_Jackiewicz}.
We analyze the Sun-v100 simulation, which features an initially uniform vertical magnetic field of 100\,G introduced into a pre-relaxed 3D hydrodynamic model.
%\vig{\sout{hydromagnetic} hydrodynamic} model.
The domain spans 38.4\,$\times$\,38.4\,$\times$\,2.8\,Mm$^{3}$ resolved on a 480\,$\times$\,480\,$\times$\,120 grid.
% While it extends to 1.3\,Mm above the optical surface, we only analyze data for 400\,km to 500\,km above the optical surface to simulate our upper photospheric IBIS diagnostics.
\added{While the model extends to a geometrical height of 1.3\,Mm above the layer where the Rosseland optical depth is unity, we only analyze data corresponding to geometrical heights of 400\,km to 500\,km to simulate the upper photosphere.}
After a 1\,hr adjustment phase to incorporate the magnetic field, the simulation proceeds for 4\,hr, and the snapshots are saved with a cadence of 30\,s.
%\vig{and the snapshots are saved} with a cadence of 30\,s.

We use the data products corresponding to the vertical velocity (v$_{\rm{z}}$) and the three components of the magnetic field (B$_{\rm{x}}$, B$_{\rm{y}}$, and B$_{\rm{z}}$).
Although at a height of 400\,km, B$_{\rm{z}}$ spans from -176\,G to 909\,G, it remains predominantly positive, with fewer than 10$\%$ of pixels exhibiting negative values.
As a result, the derived inclination angles from the horizontal component of the magnetic field $\left( B_{\rm{h}} = \sqrt{ B_{\rm{x}}^2 + B_{\rm{y}}^2} \right)$ and B$_{\rm{z}}$ predominantly cluster between $0^\circ$ to $90^\circ$.

\section{Analysis and Results} \label{sec:results}

\subsection{Vertical Velocity-Velocity Phase Differences} \label{subsec:vv_phases_differences}

\begin{figure*}[hbt!]
\gridline{\fig{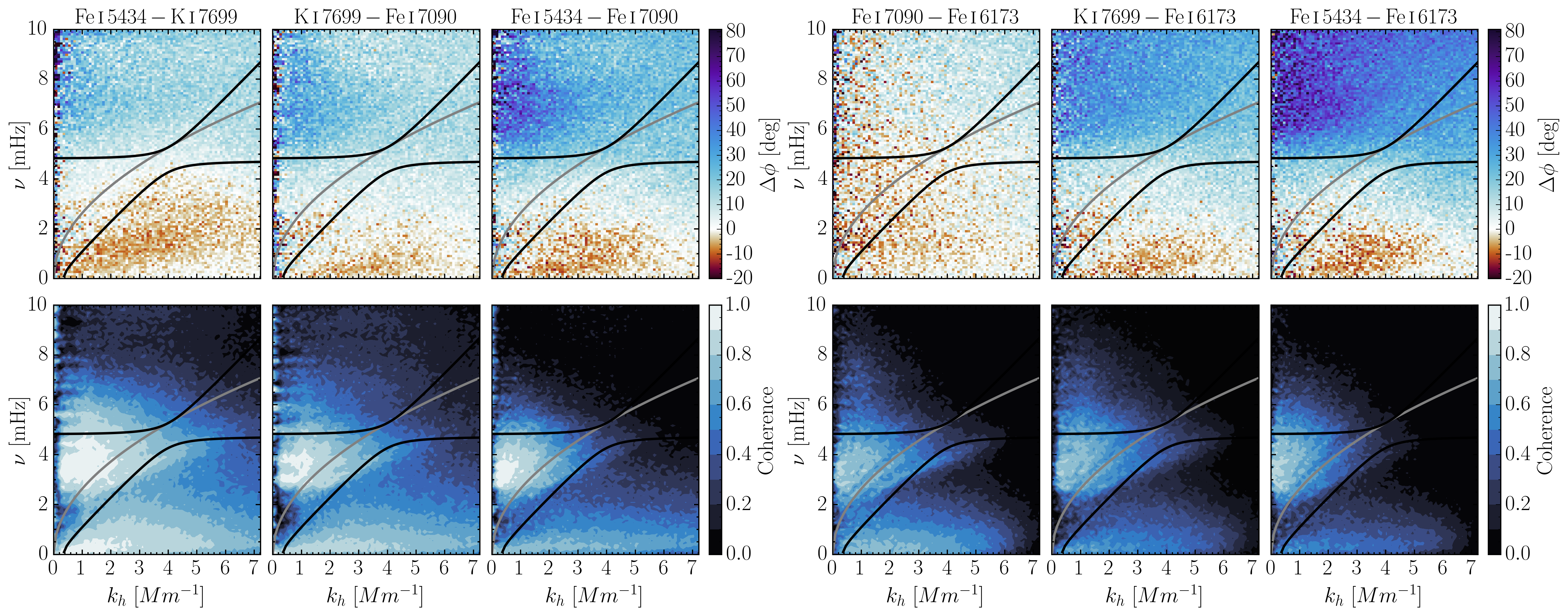}{\textwidth}{(a) DS1 ($\mu=0.57$): IBIS--IBIS \& IBIS--HMI}}
\gridline{\fig{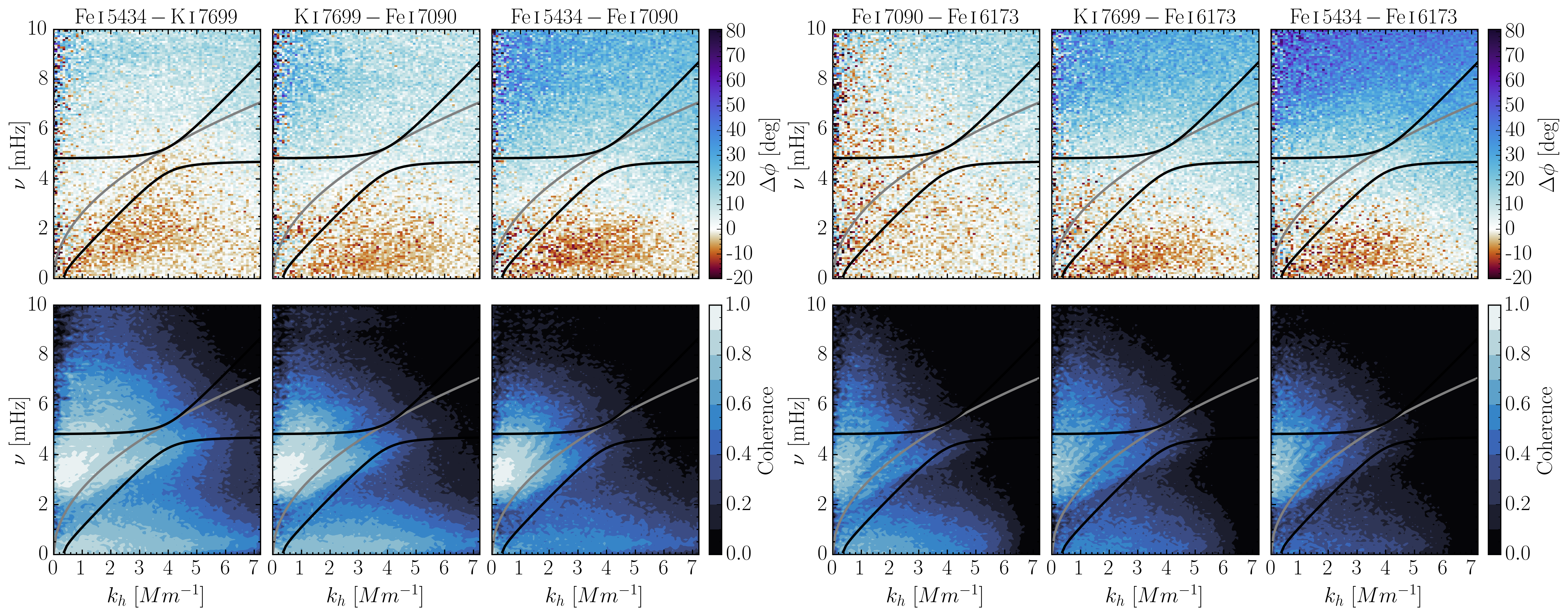}{\textwidth}{(b) DS2 ($\mu=0.64$): IBIS--IBIS \& IBIS--HMI}}
\caption{$V$\,--\,$V$ phase difference (top row) and coherence (bottom row) spectra for (left) IBIS-IBIS and (right) IBIS-HMI spectral diagnostics. In each panel, the plots are ordered by ascending separation height between diagnostics. The color scale was chosen to avoid oversaturating the vertically propagating wave regimes and to remain comparable to \citetalias{2023_Vesa}. Limiting curves for the propagating wave regimes are shown in black, and the gray curve denotes the $f$-mode.}
\end{figure*}

\begin{figure*}[hbt!]
\ContinuedFloat
\captionsetup{list=off,format=cont}
\gridline{\fig{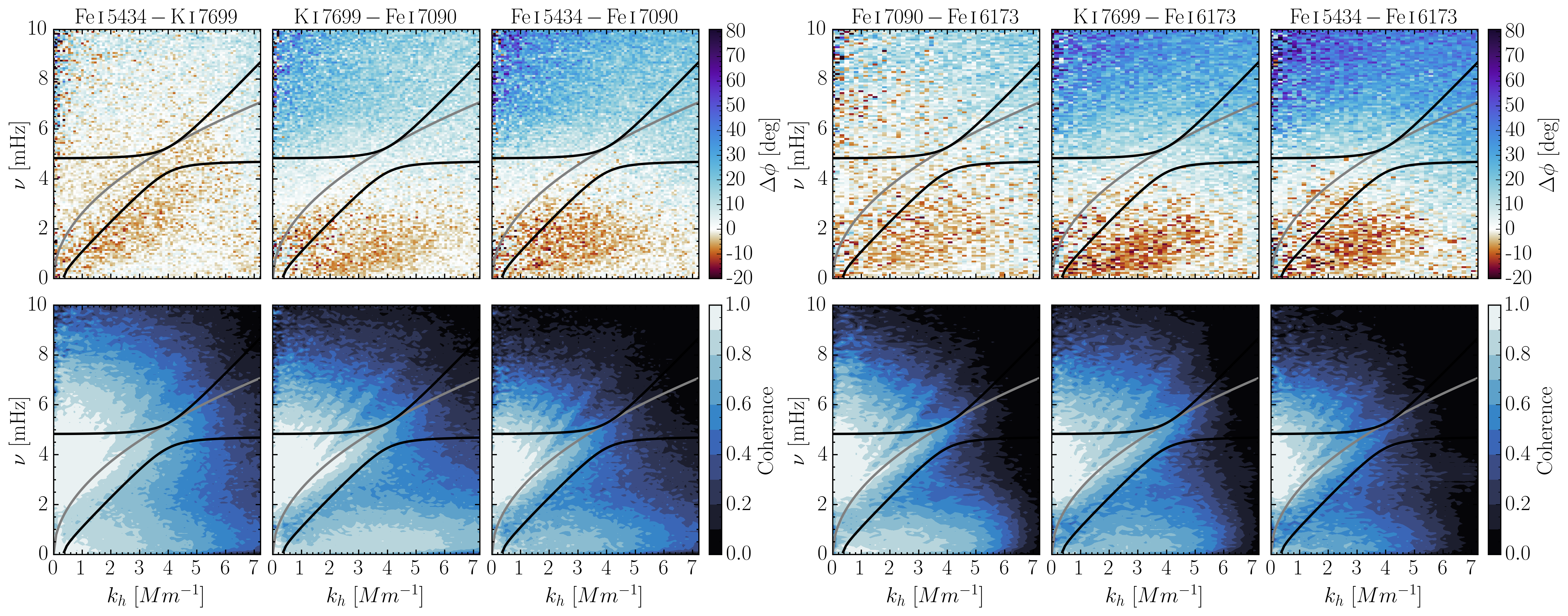}{\textwidth}{(c) DS3 ($\mu=0.88$): IBIS--IBIS \& IBIS--HMI}}
\gridline{\fig{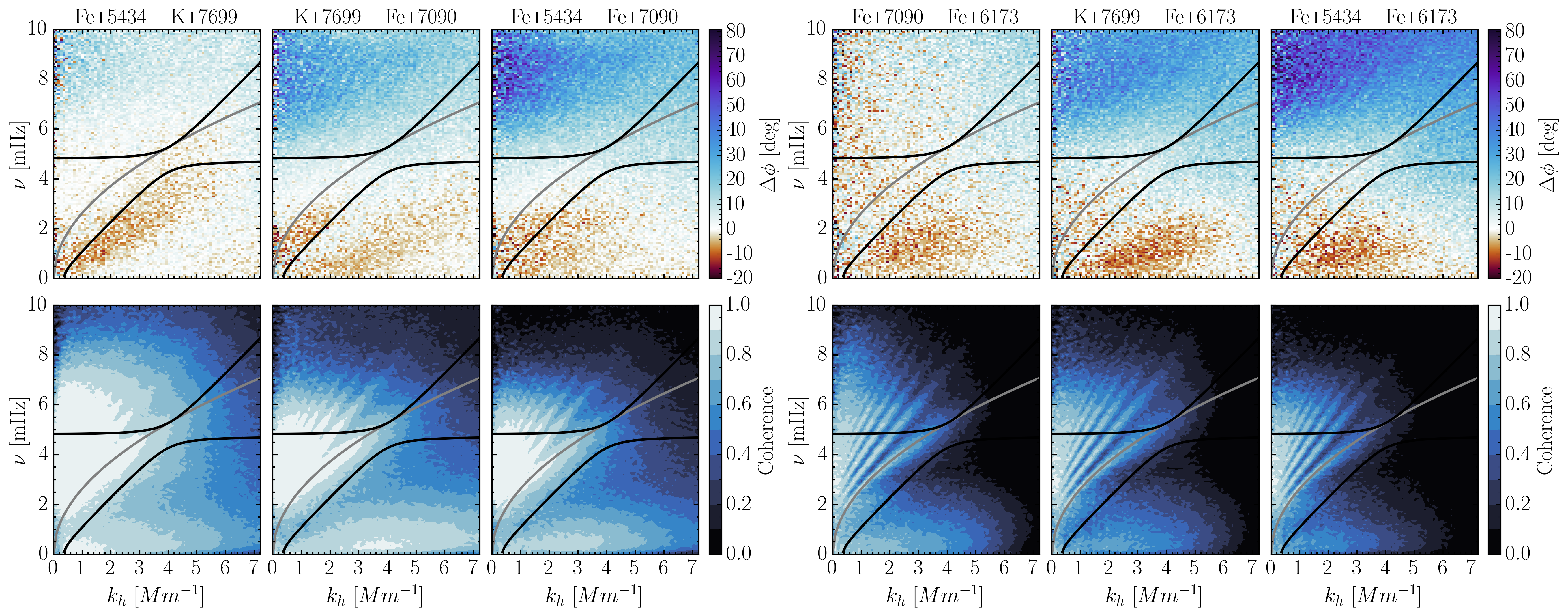}{\textwidth}{(d) DS4 ($\mu=0.95$): IBIS--IBIS \& IBIS--HMI}}
\caption{$V$\,--\,$V$ phase difference (top row) and coherence (bottom row) spectra for (left) IBIS-IBIS and (right) IBIS-HMI spectral diagnostics. In each panel, the plots are ordered by ascending separation height between diagnostics. The color scale was chosen to avoid oversaturating the vertically propagating wave regimes and to remain comparable to \citetalias{2023_Vesa}. Limiting curves for the propagating wave regimes are shown in black, and the gray curve denotes the $f$-mode.} \label{fig:vv_phase}
\end{figure*}

By applying a 3D Fourier transform to the Doppler velocity data and azimuthally averaging in the $k_{\rm{x}}-k_{\rm{y}}$ plane, we compute the phase and coherence spectra on a horizontal wavenumber-frequency ($k_{\rm{h}}-\nu$) diagram \citep{1974_Deubner} to separate the different wave behavior, as outlined in \citetalias{2023_Vesa}.
% The limiting curves are derived from Souffrin's acoustic-gravity wave dispersion equation, which describes the propagation of acoustic-gravity waves in an isothermal stratified atmosphere with constant radiative damping ($\tau_R$ = 233\,s) \citep{1966_Souffrin}. 
\added{The limiting curves are computed by solving for the acoustic-gravity wave dispersion relation detailed in \citet{1966_Souffrin} using a constant radiative damping value of $\tau_R$ = 233\,s.}
The AGW regime is in the lower left corner at low frequencies and moderate horizontal wavenumbers.

We present $V$\,--\,$V$ phase difference and coherence spectra, showing the vertical velocity perturbations, or phase lags, induced by propagating AGWs between the IBIS and HMI velocity diagnostics in Fig.\,\ref{fig:vv_phase}.
The observed phase differences agree well with simulated results \citep[e.g.,][]{2017_Vigeesh}, Souffrin's acoustic-gravity wave theory \citep[e.g.,][]{1966_Souffrin}, and wave polarization properties \citep[e.g.,][]{1981_Mihalas_Toomre, 1982_Mihalas_Toomre}.
While not shown, we find minimal coherence between combinations with \IBISCa{}.

We detect the signature of propagating AGWs carrying energy upward (negative phase differences; $\lesssim$\,-$20^\circ$) throughout the lower solar atmosphere with corresponding coherence values greater than 0.5, regardless of the viewing geometry and magnetic field configuration.
Remarkably, the phase differences between the IBIS diagnostics in Figs.\,\ref{fig:vv_phase}c and \,\ref{fig:vv_phase}d strongly resemble DS0 in \citetalias{2023_Vesa} despite the varying magnetic field \added{configuration}. 
Specifically, we want to highlight the phase distribution between \IBISFeUpperPhot{}--\IBISK{}, the highest forming velocity signals in our observations (see Subsubsect.\,\ref{subsubsec:estimated_formation_heights}).
% In both cases, these combinations showcase high-frequency AGWs while low-frequency AGWs are effectively suppressed, similarly to DS0.
\added{In both cases, these combinations showcase high-frequency AGWs while low-frequency AGWs are effectively suppressed, similarly to DS0 seen in Fig.\,4 of \citetalias{2023_Vesa}.}
The distribution has a cone-like appearance, fanning out to low frequencies and smaller horizontal wavenumbers towards the limb (DS1) and with increasing average magnetic field strength (DS2).

We also note several trends in the coherence spectra: (1) there is high coherence in the AGW regime, particularly for \IBISFeUpperPhot{}--\IBISK{}; (2) the coherence decreases with increasing separation height among diagnostics (e.g., Fig.\,\ref{fig:vv_phase}d); (3) the coherence decreases in environments with strong magnetic fields (e.g., Fig.\,\ref{fig:vv_phase}b); and (4) the coherence decreases towards the limb (e.g., compare Fig.\,\ref{fig:vv_phase}a and Fig.\,\ref{fig:vv_phase}d).

\subsubsection{Estimated Separation in Formation Heights}  \label{subsubsec:estimated_formation_heights}
\begin{deluxetable}{lr@{\,--\,}lr@{\,$\pm$\,}lr@{\,$\pm$\,}l}[htb!]
\tabletypesize{\scriptsize}
\tablewidth{0pt}
\tablecaption{Estimated formation height differences for velocity diagnostics \label{tab:estimated_formation_height_table}}
\tablehead{\colhead{Dataset} & \multicolumn{2}{c}{Line Pair} & \multicolumn{2}{c}{$\Delta\,\phi\,[\deg]$} & \multicolumn{2}{c}{$\Delta\,z\,[\textrm{km}]$}}
\startdata 
{\phantom{IBIS}} & Fe\,{\sc i}\,7090 & HMI & 4.2 & 7.3 & 17.3 & 32.3  \\
{\phantom{IBIS}} & Fe\,{\sc i}\,5434 &  K\,{\sc i}\,7699 & 17.7 & 6.5 & 77.8 & 35.5 \\
{DS1} &  K\,{\sc i}\,7699 & Fe\,{\sc i}\,7090  & 21.8 & 7.7 & 97.9 & 48.4\\
{\phantom{IBIS}} & K\,{\sc i}\,7699 & HMI & 34.3 & 7.2 & 149.1 & 51.0\\
{\phantom{IBIS}} & Fe\,{\sc i}\,5434 & Fe\,{\sc i}\,7090 & 35.1 & 10.3 & 157.3 & 71.9  \\
{\phantom{IBIS}} & Fe\,{\sc i}\,5434 & HMI & 50.7 & 8.6 & 221.7 & 74.1\\
\hline
{\phantom{IBIS}} & Fe\,{\sc i}\,7090 & HMI & 5.2 & 7.1 & 21.5 & 31.8  \\
{\phantom{IBIS}} & Fe\,{\sc i}\,5434 &  K\,{\sc i}\,7699 & 8.4 & 5.9 & 33.9 & 23.0 \\
{DS2} &  K\,{\sc i}\,7699 & Fe\,{\sc i}\,7090  & 14.3 & 7.1 & 60.0 & 29.4\\
{\phantom{IBIS}} & K\,{\sc i}\,7699 & HMI & 23.1 & 8.3 & 95.4 & 31.5\\
{\phantom{IBIS}} & Fe\,{\sc i}\,5434 & Fe\,{\sc i}\,7090 & 24.3 & 8.6 & 101.4 & 34.2\\
{\phantom{IBIS}} & Fe\,{\sc i}\,5434 & HMI & 33.5 & 10.1 & 138.3 & 33.5\\
\hline
{\phantom{IBIS}} & Fe\,{\sc i}\,5434 &  K\,{\sc i}\,7699 & 2.8 & 4.6 & 11.6 & 18.7 \\
{\phantom{IBIS}} & Fe\,{\sc i}\,7090 & HMI & 8.7 & 8.0 & 35.6 & 34.2  \\
{DS3} &  K\,{\sc i}\,7699 & Fe\,{\sc i}\,7090  & 19.4 & 6.6 & 82.6 & 29.7\\
{\phantom{IBIS}} & Fe\,{\sc i}\,5434 & Fe\,{\sc i}\,7090  & 24.5 & 8.2 & 103.2 & 3.4\\
{\phantom{IBIS}} & K\,{\sc i}\,7699 & HMI & 25.9 & 8.9 & 108.0 & 36.8\\
{\phantom{IBIS}} & Fe\,{\sc i}\,5434 & HMI & 31.5 & 9.5 & 131.5 & 37.3\\
\hline
{\phantom{IBIS}} & Fe\,{\sc i}\,7090 & HMI & 3.5 & 5.7 & 14.3 & 25.0  \\
{\phantom{IBIS}} & Fe\,{\sc i}\,5434 &  K\,{\sc i}\,7699 & 6.8 & 4.8 & 26.3 & 16.0 \\
{DS4} &  K\,{\sc i}\,7699 & Fe\,{\sc i}\,7090  & 24.4 & 7.4 & 102.6 & 29.8\\
{\phantom{IBIS}} & K\,{\sc i}\,7699 & HMI & 30.4 & 8.6 & 126.5 & 30.1\\
{\phantom{IBIS}} & Fe\,{\sc i}\,5434 & Fe\,{\sc i}\,7090  & 32.8 & 10.2 & 134.5 & 30.7\\
{\phantom{IBIS}} & Fe\,{\sc i}\,5434 & HMI & 40.0 & 12.2 & 163.1 & 32.3\\
\enddata
\tablecomments{The measured $\Delta\,\phi$ are averages found in the propagating acoustic wave regime above the acoustic cutoff frequency within the range of 6–9\,mHz and 1–3\,Mm$^{-1}$. We assume nominal photospheric values for the computation of $\Delta$\,z, such as an acoustic cutoff frequency of 5.4\,mHz, a Brunt–V\"{a}is\"{a}l\"{a} frequency of 4.9\,mHz, and a photospheric sound speed of 7.0\,km\,s$^{-1}$. The total phase speed for this frequency and horizontal wavenumber range for all datasets is 11.5\,$\pm$\,2.4\,kms$^{-1}$.}
\end{deluxetable}

The average ``formation height'' of spectral lines is contingent upon atmospheric properties and viewing geometries.
Commonly quoted values in the literature, including those in Table\,\ref{tab:dataset_and_spectral_line_properties}, are estimates based on the atmospheric conditions present at QS disk center where the line core velocity signal samples.
However, the properties of spectral lines (i.e., the overall shape of the profile and average formation height) differ even at QS disk center \citep[for instance, granular vs. intergranular regimes;][]{2001_Shchukina_TrujilloBueno, 2011_kneer_Bello}, are subject to viewing geometries or center-to-limb variations \citep[e.g.,][]{1975_Altrock_November_Simon_etal, 2019_LohnerBottcher_Schmidt_Schlichenmaier_etal}, and are affected by the temperature and magnetic field strength \citep[i.e., flares, ARs, sunspot umbra;][]{2006_Norton_Graham_Ulrich_etal, 2017_QuinteroNoda, 2020_Kuridze_Mathioudakis_Heinzel_etal}.

Using the method outlined in \citetalias{2023_Vesa}, we measure the estimated separation in formation height ($\Delta$\,z) between spectral diagnostics using the observed phase differences ($\Delta$\,$\phi$) detected in the propagating acoustic wave regime within the range 6--9\,mHz and 1--3\,Mm$^{-1}$.
The results are shown in Table\,\ref{tab:estimated_formation_height_table}.

Analysis of $\Delta$\,z between diagnostics elucidates several key findings: (1) all datasets display the same relative order for the formation heights of IBIS and HMI diagnostics; (2) $\Delta$\,z and $\Delta$\,$\phi$ decrease in magnitude towards the limb; and (3) $\Delta$\,z and $\Delta$\,$\phi$ are smaller in magnitude in ARs than in QS regions.

Our analysis shows that, irrespective of the magnetic field and viewing geometry, the relative formation order of the IBIS and HMI Dopplergram diagnostics remains comparable, as seen in Fig.\,\ref{fig:vv_phase}.
The decreasing $\Delta$\,z values towards the limb and in regions with large magnetic fields imply that collectively the average formation height of the diagnostics varies compared to QS disk center, which agrees with previous independent investigations \citep{2012_Faurobert_Ricort_Aime, 2017_QuinteroNoda}.
\added{Toward the limb, we expect to observe higher in the atmosphere, with the average formation heights of spectral lines increasing compared to disk center \citep{2012_Faurobert_Ricort_Aime}.}
% Toward the limb, the line-of-sight velocities are expected to have a greater horizontal component; thus, we would expect to observe deeper into the atmosphere, which would shift up the average formation heights \citep{2012_Faurobert_Ricort_Aime}.
In regions with large magnetic fields, we would expect the average formation heights of the diagnostics to decrease compared to their QS counterparts \citep{2003_Uitenbroek, 2017_QuinteroNoda, 2020_Vigeesh_Roth}, which in fact can be significantly smaller \citep[e.g., Fig.\,9 in][]{2017_QuinteroNoda}.
This is reflected in our results, as seen in Table\,\ref{tab:estimated_formation_height_table}.
DS2 has the largest average magnetic field (see Table\,\ref{tab:magnetic_field_information}) but some of the smallest $\Delta$\,z values, indicating that this magnetic environment impacted the formation heights of our diagnostics the most. 
We also note that there is no apparent linear correlation between the estimated $\Delta$\,z values seen at DS0 and those reported at the QS limb (DS1; $\mu$ = 0.57) or the AR limb (DS2; $\mu$ = 0.64).

\subsubsection{Magnetic Dispersion Relations}
\label{subsubsec:magnetic_dispersions}

To facilitate comparisons between the observed and theoretical $k_{\rm{h}}-\nu$ phase difference diagrams, we will use a more physically motivated dispersion relation for AGWs.
The standard dispersion relations used to compare observations to theory are those of 
\citet{1966_Souffrin, 1981_Mihalas_Toomre, 1982_Mihalas_Toomre}. These models describe the propagation of acoustic-gravity waves in an isothermal, vertically stratified medium, and in the case of \citet{1966_Souffrin, 1982_Mihalas_Toomre}, one with radiative damping. However, these models do not account for the effects of magnetic fields. Such effects are taken into account in the models of \citet{2006_Schunker_Cally, 2009_Newington_Cally}. The authors find that magnetic fields and, more importantly, inclined fields have significant effects on the behavior of AGWs that can depart drastically from the simplified theory. 

\begin{figure*}[hbt!]
\gridline{\fig{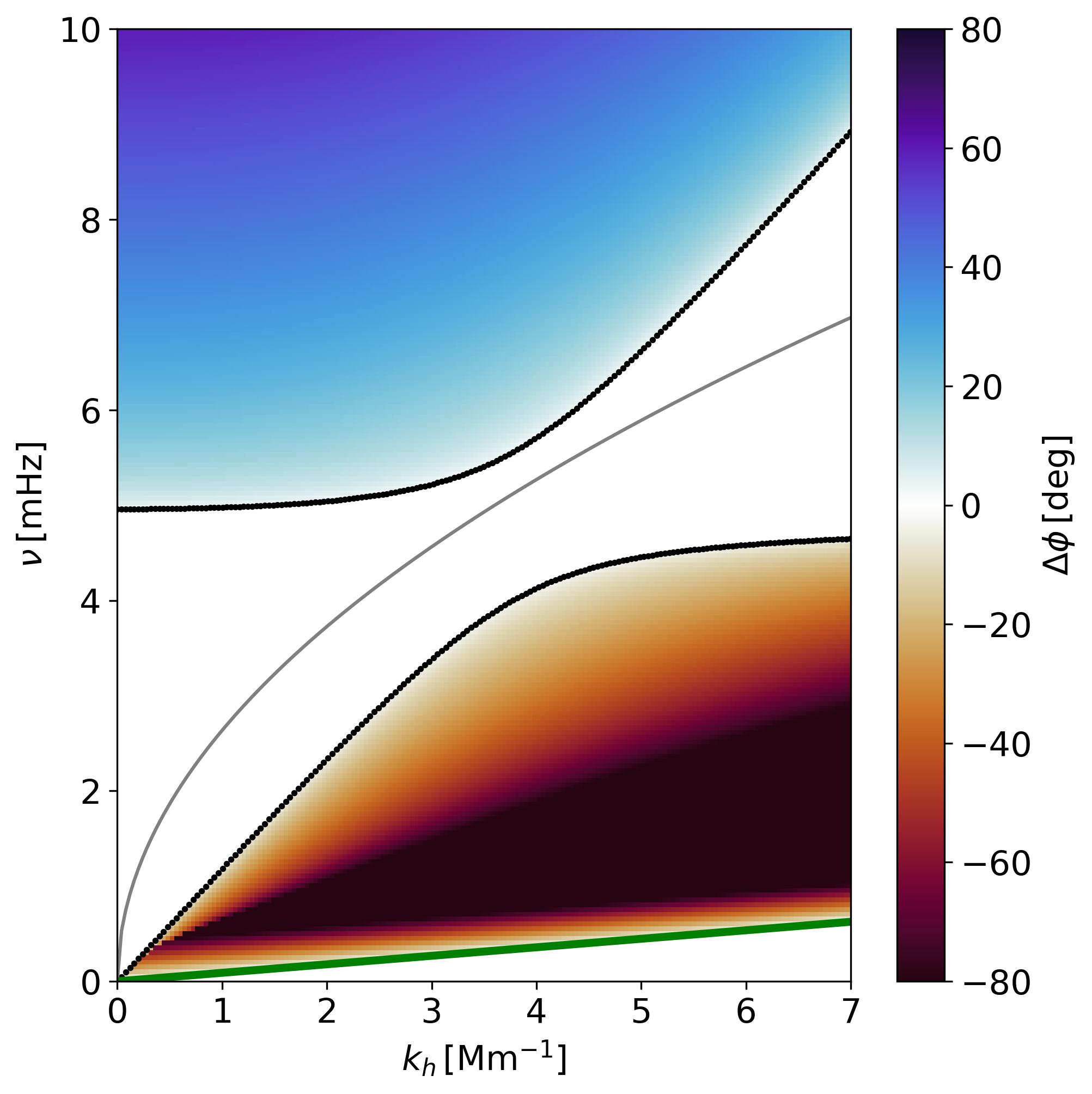}{0.3\textwidth}{(a) $a = 0.10c_s$; $\theta = 70^{\circ}$; $\gamma = 50^{\circ}$.}
          \fig{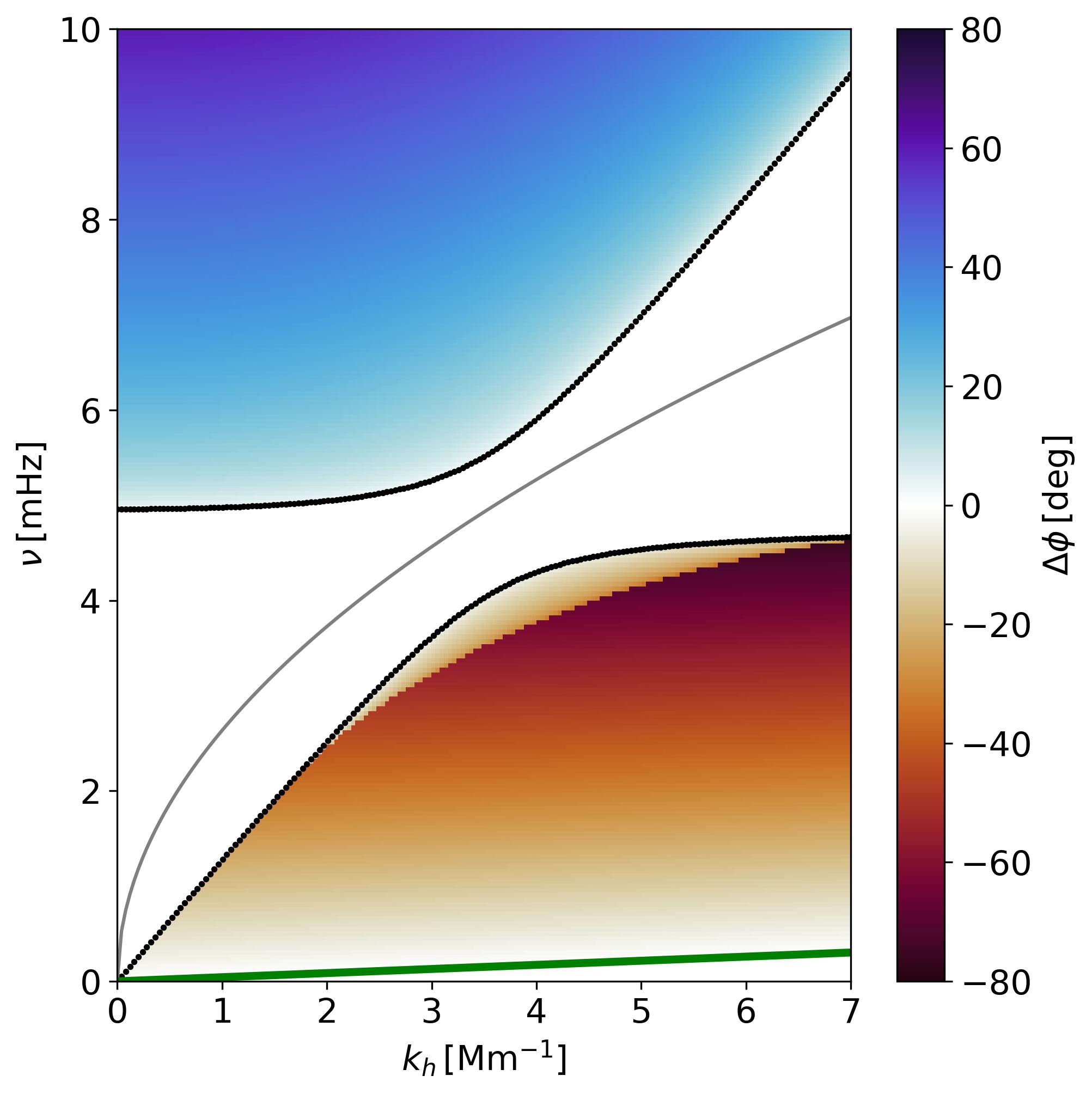}{0.3\textwidth}{(b) $a = 0.40c_s$; $\theta = 10^{\circ}$; $\gamma = 30^{\circ}$}
          \fig{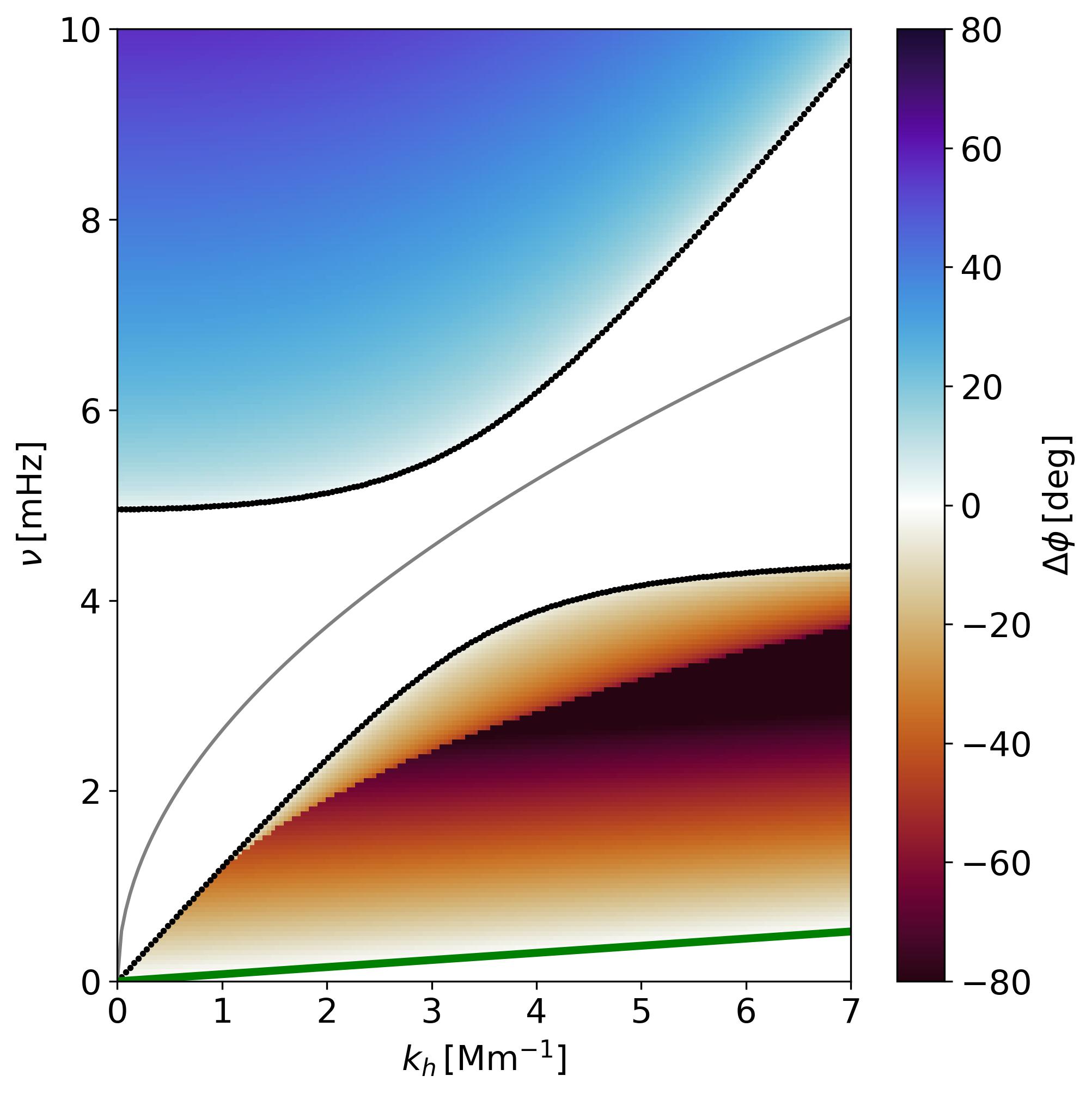}{0.3\textwidth}{(c) $a = 0.40c_s$; $\theta = 60^{\circ}$; $\gamma = 10^{\circ}$}}
\gridline{\fig{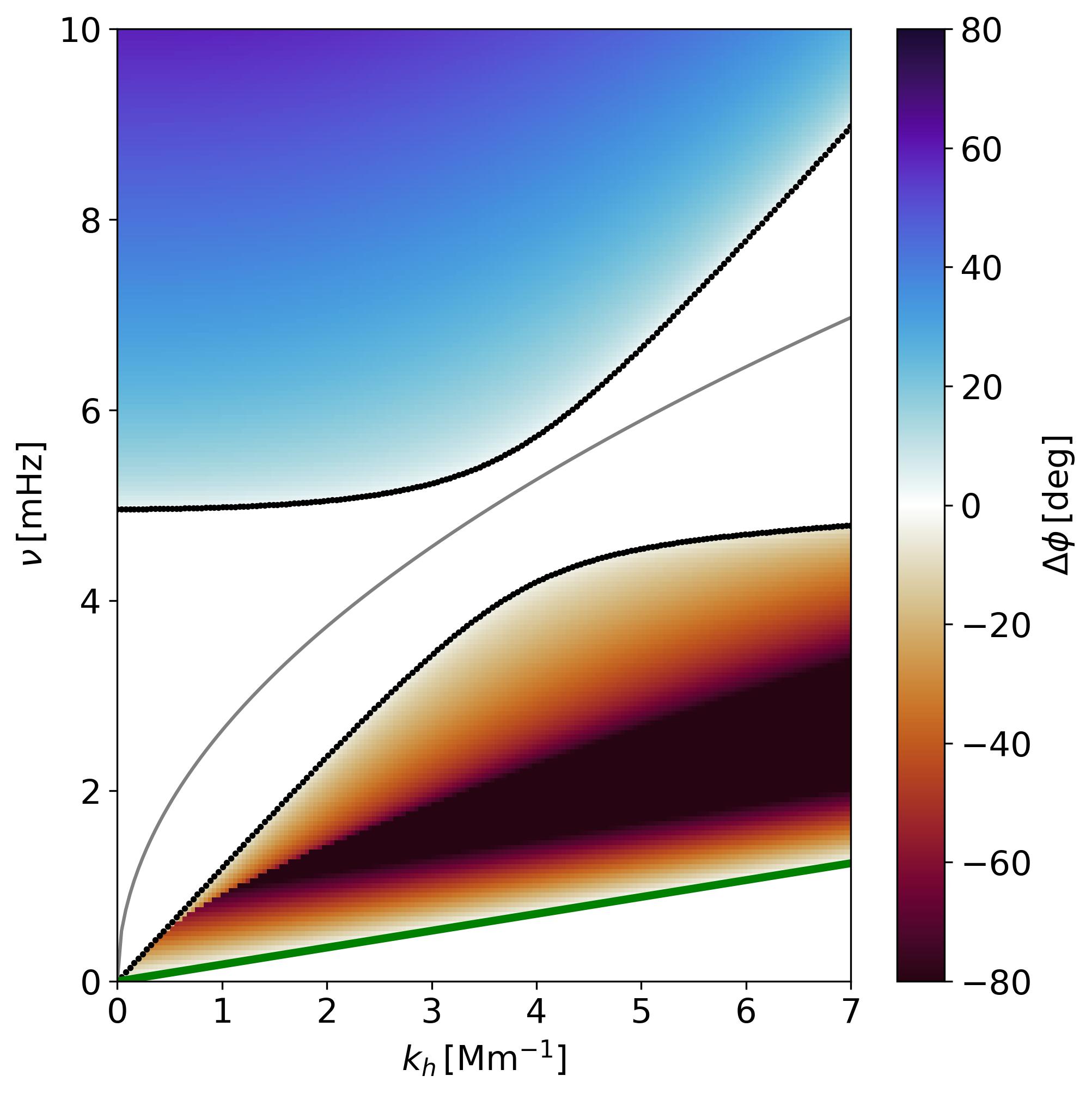}{0.3\textwidth}{(d) $a = 0.20c_s$; $\theta = 70^{\circ}$; $\gamma = 50^{\circ}$}
          \fig{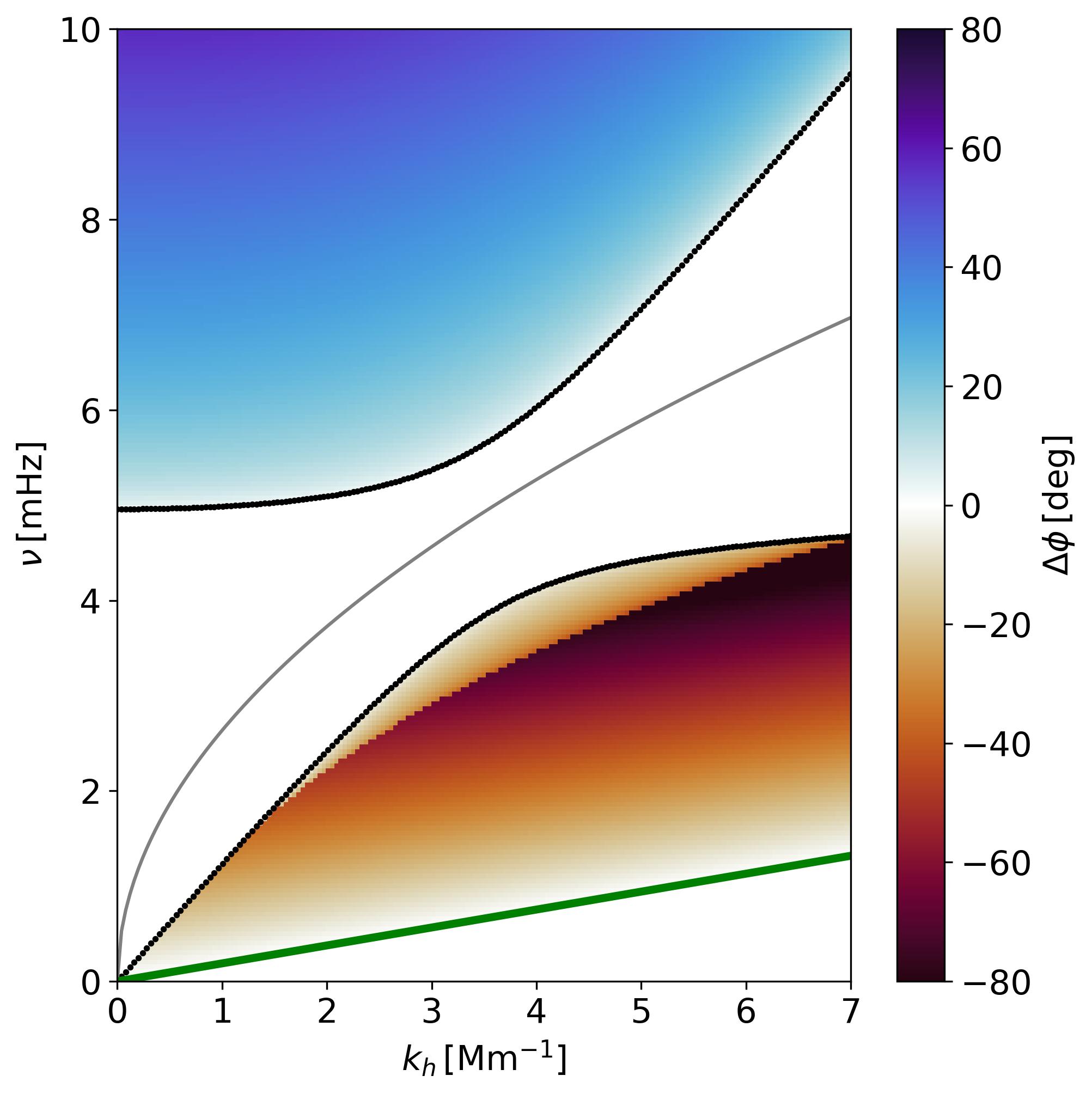}{0.3\textwidth}{(e) $a = 0.40c_s$; $\theta = 50^{\circ}$; $\gamma = 30^{\circ}$}
          \fig{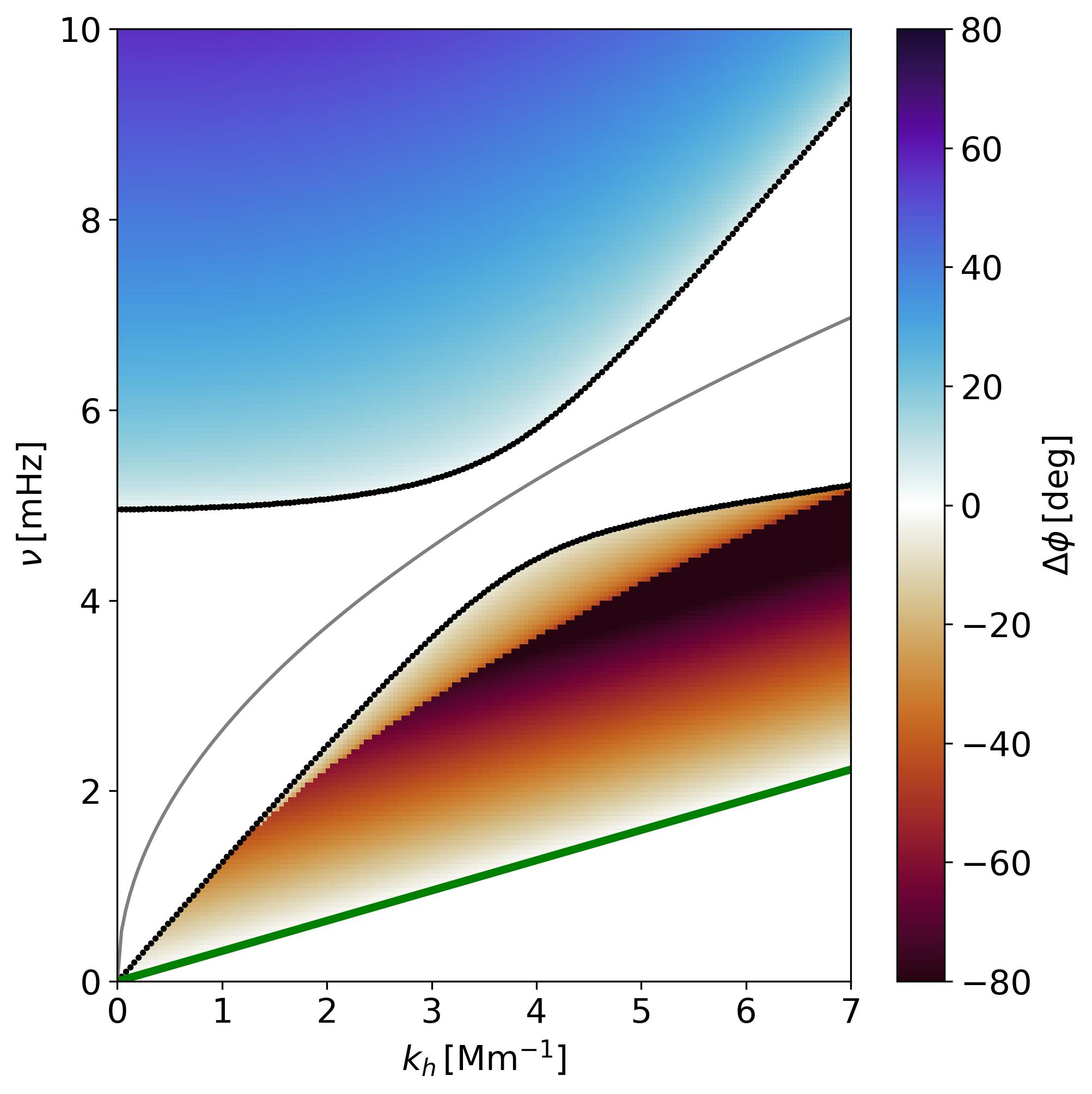}{0.3\textwidth}{(f) $a = 0.40c_s$; $\theta = 60^{\circ}$; $\gamma = 50^{\circ}$}}
\gridline{\fig{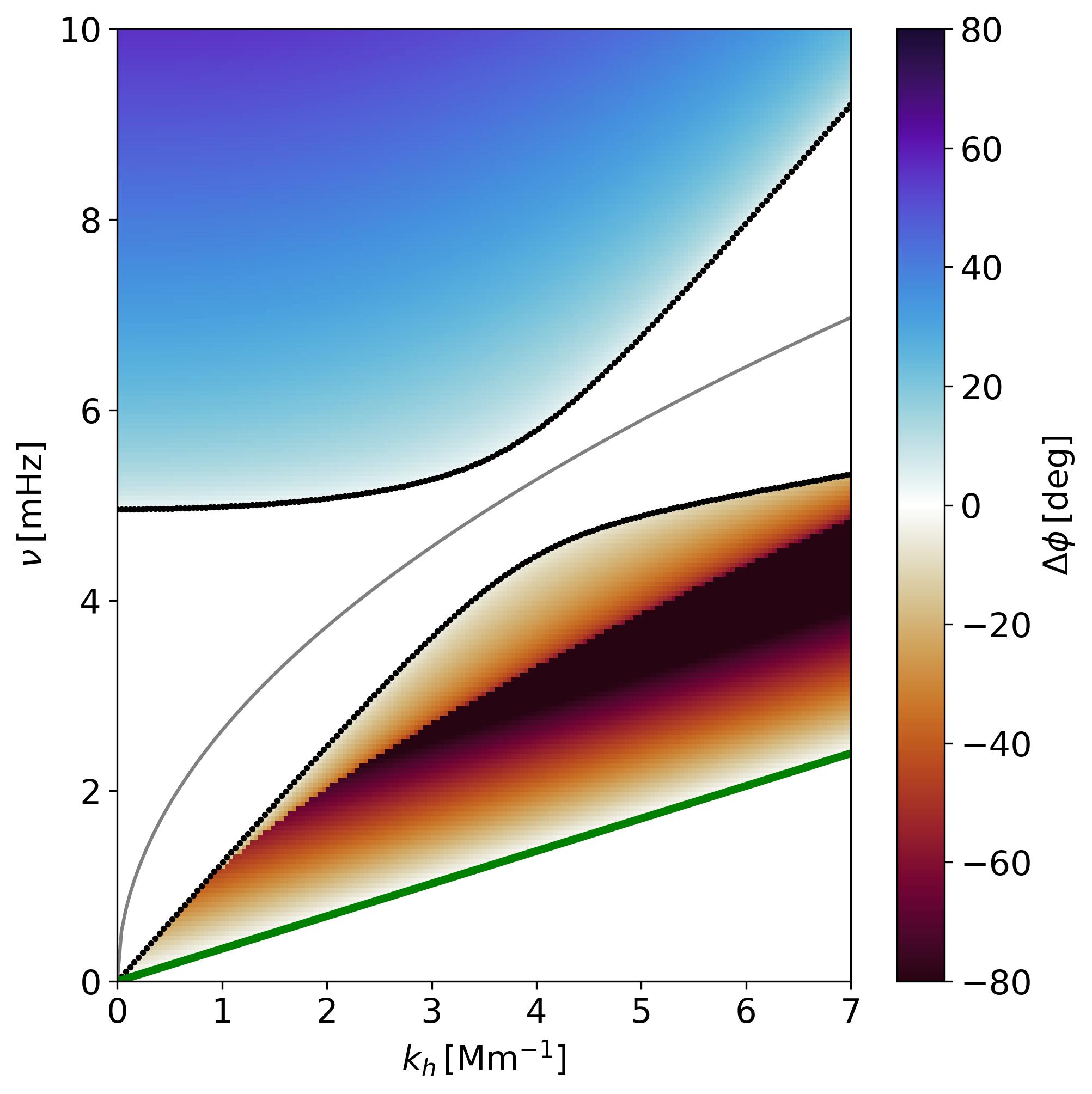}{0.3\textwidth}{(g) $a = 0.40c_s$; $\theta = 70^{\circ}$; $\gamma = 50^{\circ}$}
          \fig{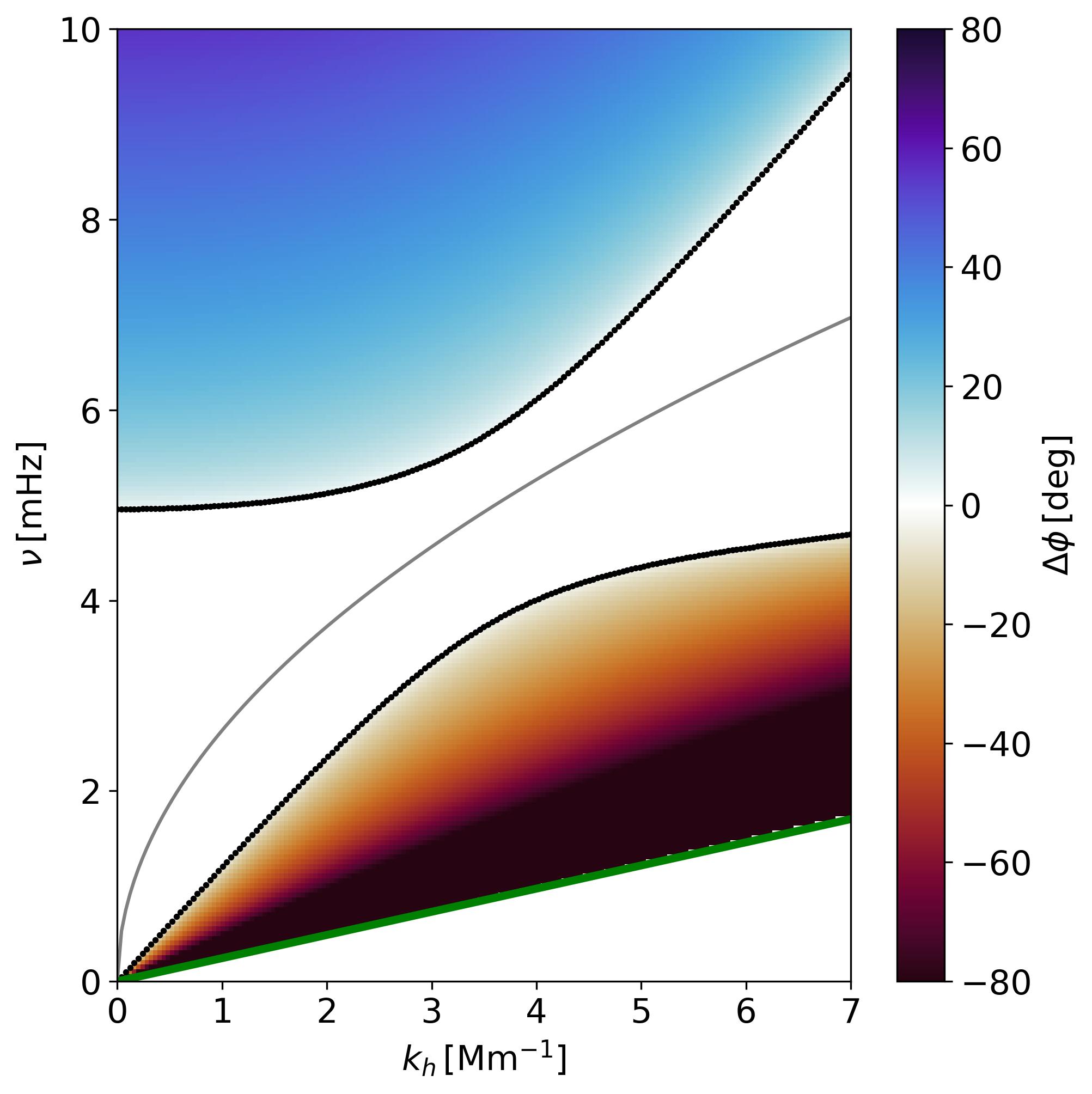}{0.3\textwidth}{(h) $a = 0.40c_s$; $\theta = 90^{\circ}$; $\gamma = 30^{\circ}$}
          \fig{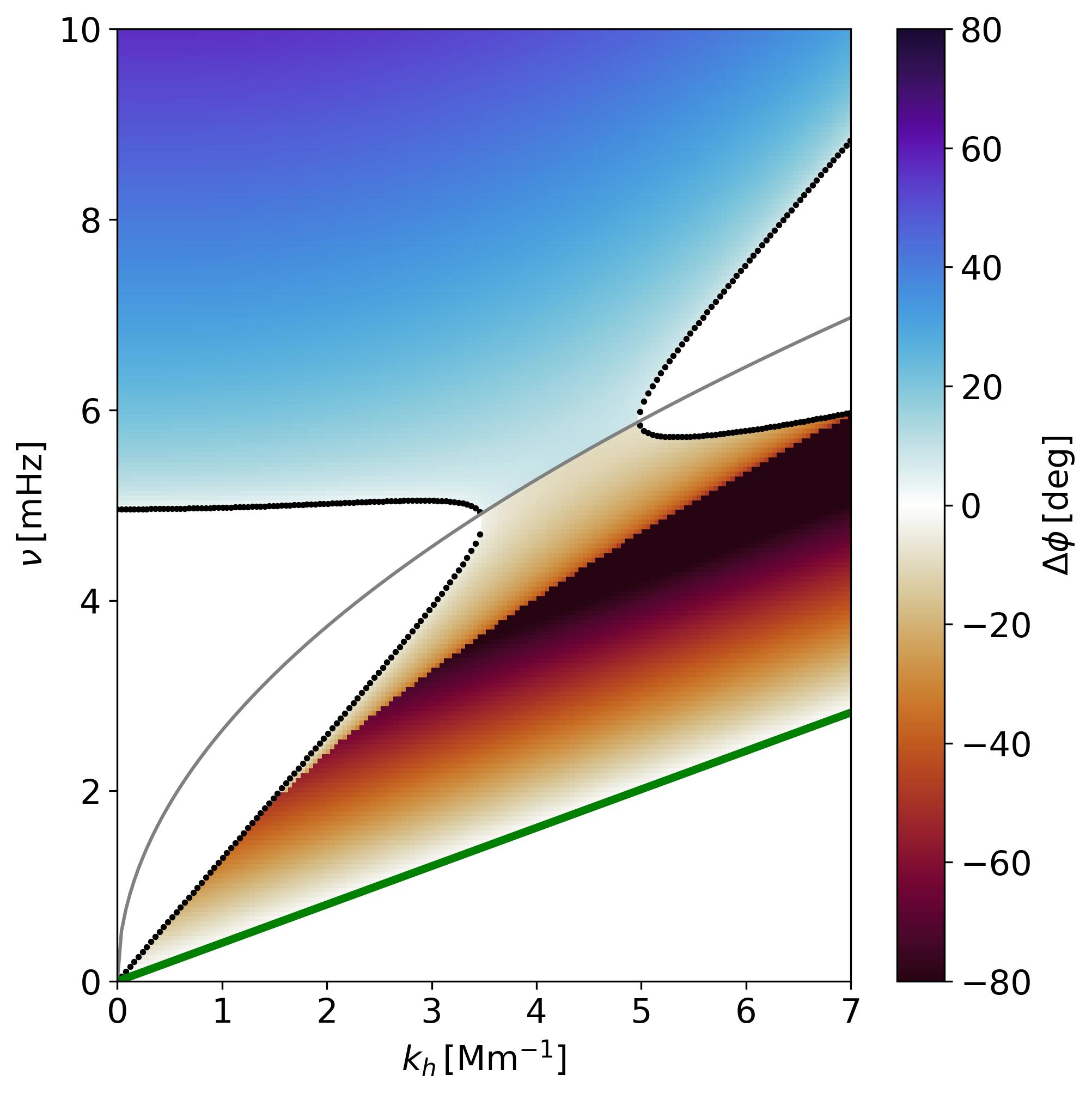}{0.3\textwidth}{(i) $a = 0.40c_s$; $\theta = 60^{\circ}$; $\gamma = 90^{\circ}$}}
\caption{Simulated $k_{\rm{h}}-\nu$ phase difference diagrams for various parameter sets, calculated using the 3D MHD dispersion relation from \citet{2009_Newington_Cally}. The leftmost column shows the effects of varying the Alfvén speed ($a$) while keeping the field inclination angle ($\theta$) and azimuth ($\gamma$) fixed. The middle column varies $\theta$ while holding $a$ and $\gamma$ constant, and the rightmost column varies $\gamma$ with $a$ and $\theta$ held constant. Black curves denote the $k_z^2 < 0$ boundaries in a magnetized medium, and the light gray curve represents the $f$-mode dispersion relation. The green line is the lower boundary of the AGW regime and appears to be correlated with field properties. All diagrams assume a height separation of $\Delta\,z = 150\,\mathrm{km}$.
} \label{fig:mag_diagnostic_diagrams}
\end{figure*}

When compared with observations, such as those in Fig.\,\ref{fig:vv_phase}, the AGW regime is found to be poorly constrained by the simplistic models. To address this, we start with the full 3D magnetohydrodynamic dispersion relation from \citet{2009_Newington_Cally}:
\begin{equation}
    \begin{split}
        \mathcal{D} =\; & \omega^2 \omega_c^2 a_y^2 k_h^2 
        + \left(\omega^2 - a^2 k_{\parallel}^2\right) \times \\
        & \left[
        \omega^4 
        - (a^2 + c^2)\omega^2 k^2 
        + a^2 c^2 k^2 k_{\parallel}^2 
        + c^2 N^2 k_h^2 \right. \\
        & \left. \quad
        - \left(\omega^2 - a_z^2 k^2\right)\omega_c^2
        \right],
    \end{split}
\label{eq:3dmhd_dispersion}
\end{equation}
where $\omega$ is the angular frequency, $k$ is the total wavenumber where $k=\sqrt{k_h^2 + k_z^2}$, $a$ is the Alfv\'{e}n speed where $a \propto B$ and $B$ is the magnetic field strength, $a_y$ is its component perpendicular to the vertical ($x-z$) plane, $N$ is the Brunt–Väisälä frequency, $\omega_c$ is the acoustic cut-off frequency, $c_s$ is the sound speed, and $k_{\parallel}$ is the component of the wavenumber vector parallel to the magnetic field. $\mathcal{D} = 0$ in Eq.\,\ref{eq:3dmhd_dispersion} limit solutions to $k_{\rm{h}}-\nu$ space. To create a magnetized version of the phase difference diagram, Eq.\,\ref{eq:3dmhd_dispersion} must be solved explicitly for $k_z$. Expanding all terms in Eq.\,\ref{eq:3dmhd_dispersion} and rearranging, one finds the dispersion relation can be expressed in the form of a 6th order polynomial in $k_z$:
\begin{equation}
    c_6 k_z^6 + c_5 k_z^5 + c_4 k_z^4 + c_3 k_z^3 + c_2 k_z^2 + c_1 k_z + c_0  = 0,
    \label{eq:6th_order_polynomial}
\end{equation}
where the coefficients $c_i$ are functions of the other parameters in the dispersion relation, most crucially $a\text{, }\theta\text{, and } \gamma$. $\theta$ is the inclination angle of the magnetic field with respect to the $z$-axis, and $\gamma$ is the field azimuth. This equation can be solved numerically for the roots of $k_z$. Then the phase difference between the two heights is given by:
\begin{equation}
    \Delta \phi = k_z \Delta z.
    %\Delta \phi = \frac{2\pi \nu \Delta z}{v_{p, z}},
    \label{fig:phase_difference}
\end{equation}
%where $v_{p, z} = \omega/k_z$ is the phase speed in the vertical direction and $\nu = \omega/2\pi$. 
Figure\,\ref{fig:mag_diagnostic_diagrams} shows select phase difference diagrams calculated from Eq.\,\ref{fig:phase_difference} with varying values of the parameters $a\text{, }\theta\text{, and } \gamma$. In all examples, the phase is found to be highly suppressed in the AGW regime. Immediately, we see remarkable qualitative similarities between the magnetized diagrams and a few observations from Fig.\,\ref{fig:vv_phase} and \citetalias{2023_Vesa}, particularly those of \IBISFeUpperPhot{}\,--\,\IBISK{}. Yet, the phase difference diagrams fail to reproduce most of the $k_{\rm{h}}-\nu$ diagrams in Fig.\,\ref{fig:vv_phase}, highlighting the limited discernibility of the effects of the magnetic field on these diagrams (to be discussed further in Sec.\,\ref{sec:discussion}).

% The slope of the lower boundary of the AGW regime is found to be generally correlated to the Alfv\'{e}n speed and therefore the magnetic field strength. We can derive an analytic expression for the slope of the boundary, assuming a linear form
% \begin{equation}
%     \omega = A k_h + B.
%    \label{eqn:lower_boundary}
% \end{equation}
% By visual inspection of Fig.\,\ref{fig:mag_diagnostic_diagrams}, we see that $B \approx 0$, thus we only need to solve for the coefficient $A$. We do this by substituting the linear form into Eq.\,\ref{eq:6th_order_polynomial} and solving for $A$. Since we are interested in the boundary $k_z=0$, Eq.\,\ref{eq:6th_order_polynomial} becomes

% \begin{equation}
%     c_0(a, \theta, \gamma) = 0.
%     \label{eq:6th_order_polynomial_boundary}
% \end{equation}

% Eq.\,\ref{eq:6th_order_polynomial_boundary} has an analytic solution for $A$ in terms of $a\text{, }\theta\text{, and } \gamma$. The expression is too complex to include here, but we find that it accurately fits the slope of the lower boundary in Fig.\,\ref{fig:mag_diagnostic_diagrams} shown as the green line. 
The slope of the lower boundary of the AGW regime at low frequencies (shown in green) is found to be generally correlated to the Alfv\'{e}n speed and the magnetic field orientation.
If the wedges of suppressed AGWs at low frequencies in the observations are indeed due to the presence of magnetic fields, then we may be able to use the slope of the lower boundary to crudely infer the spatiotemporal average field properties in the lower solar atmosphere. This hypothesis will be validated using the 3D MHD simulations of \citet{2017_Vigeesh, 2020_Vigeesh_Roth, 2021_Vigeesh_Roth_Steiner_Fleck} in upcoming work.

%represents a linear analytic solution to $k_z^2 = 0$ in terms of the magnetic field properties. 
\subsection{Spatial Coherence-weighted Phase Difference Maps} \label{sec:spatial_distribution_phases}
\begin{figure*}
    \centering
    \includegraphics[width=\linewidth]{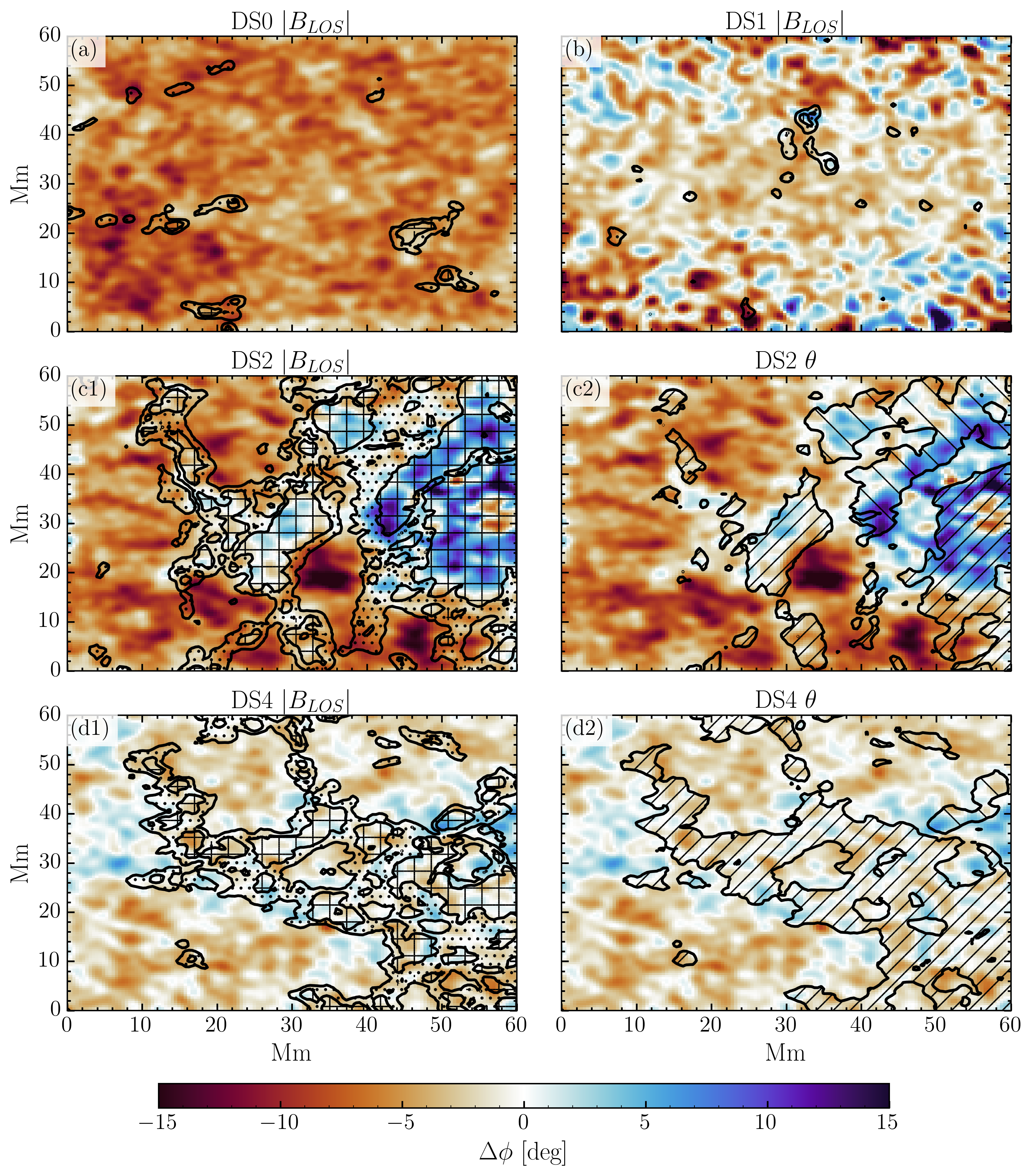}
    \caption{Spatial coherence-weighted phase difference between the filtered \IBISFeUpperPhot{}--\IBISK{} diagnostics, integrated within 0.7--3.1\,mHz, overlaid with the temporally averaged unsigned magnetic field and inclination contours. Regions outside the contours correspond to QS ($|B| < 30$\,G) and predominantly horizontal fields ($75\degree \leq \theta \leq 106\degree$). Inclination data are unavailable for DS0 and DS1. Magnetic field contours: dotted hatching indicates intermediate fields ($30$\,G $< |B| \leq 100$\,G); square hatching indicates strong fields ($|B| > 100$\,G). Inclination contours: vertically oriented fields are hatched (upward oriented fields are marked with right-leaning diagonals: $0\degree \leq \theta < 75\degree$; downward oriented fields are marked with left-leaning diagonals: $106\degree < \theta \leq 180\degree$).}
    \label{fig:IBIS_spatial_distribution_phases}
\end{figure*}

It is important to note that $k_{\rm{h}}-\nu$ phase difference diagrams effectively capture the vertical propagation of AGWs, and those simulated in Fig.\,\ref{fig:mag_diagnostic_diagrams} show sensitivity to field properties.
However, we detect minimal observational effects of the magnetic \added{field configuration} on the behavior of AGWs, unlike those reported in numerical simulations by \citet{2017_Vigeesh,2019_Vigeesh_Roth_Steiner_Jackiewicz}.
Therefore, we employ spatial coherence-weighted phase difference maps averaged over frequencies of interest.
This technique has been successfully used to uncover the effects of the magnetic field on $p$-mode oscillations \citep[e.g.,][]{2006_Jefferies_McIntosh_etal, 2007_Vecchio_Cauzzi_Reardon_etal}.

\added{However, to minimize interference from the neighboring $p$-modes at low frequencies and horizontal wavenumbers where AGWs are present, we apply a 3D Gaussian filter in frequency-wavenumber space to the velocity cubes in Fourier space, tapered well before the Lamb line (see \citetalias[][]{2023_Vesa}).}
% To avoid interference from the neighboring $p$-modes, we isolate AGWs by applying a 3D Gaussian filter in both frequency and wavenumber space to the velocity cubes in Fourier space, tapered off well before the Lamb line (see \citetalias[][]{2023_Vesa}).
The resulting filter is slightly non-Gaussian with a central frequency $\nu = 1.5$\,mHz and central horizontal wavenumber $k_{\rm{h}} = 2.0$\,Mm$^{-1}$, representing where the bulk of AGWs reside (see Fig.\,\ref{fig:vv_phase}). 

A 1D FFT was applied along the temporal axis of the filtered Doppler diagnostics, and the corresponding power, cross-spectra, phase, and coherence were computed.
The phase and coherence spectra were averaged over frequency bands associated with AGWs (0.7--3.1\,mHz) and smoothed spatially with a 1D Gaussian filter of full width at half maximum of 0.7\,Mm.
Because we are only interested in the properties of the average magnetic field, the resulting spatial coherence-weighted phase difference maps are compared with HMI's temporally averaged $|B_{LOS}|$ and field inclination, with contours overlaid for reference.
Contours of $|B_{LOS}|$ were divided into QS fields ($|B|$\,$\leq$\,30\,G), intermediate fields (30\,G\,$<$\,$|B|$\,$\leq$\,100\,G), and strong fields ($|B|$\,$>$\,100\,G), similarly to \citet{2019_Jefferies_Fleck_Murphy_Berrilli}. 
Following the classification by \citet{2021_Campbell_Mathioudakis_Collados_etal}, contours for the field inclination were divided to capture vertical ($0\degree$\,$\leq$\,$\theta$\,$<$\,$75\degree$ and $106\degree$\,$<$\,$\theta$\,$\leq$\,$180\degree$) and transverse (horizontal) fields ($75\degree$\,$\leq$\,$\theta$\,$\leq$\,$106\degree$).

\subsubsection{IBIS Observations} \label{subsubsec:spatial_ibis_observations}

The spatial coherence-weighted phase difference maps for the filtered \IBISFeUpperPhot{}--\IBISK{} diagnostics are seen in Fig.\,\ref{fig:IBIS_spatial_distribution_phases}.
We focus on the \IBISFeUpperPhot{}--\IBISK{} combination because these diagnostics sample the highest atmospheric regions, where magnetic field effects on AGW propagation are expected to be most significant \citep{2019_Vigeesh_Roth_Steiner_Jackiewicz}.
Regions outside the hatched areas represent QS fields and predominantly transverse fields.
Due to projection effects near the limb, DS1 contours primarily reflect the longitudinal component of the field rather than the radial. 

While the HMI LOS and vector magnetograms sample the low photosphere, and our spectral diagnostics probe higher regions, these measurements still offer a reasonable approximation of the magnetic environment relevant to the propagation of AGWs.
Due to the expansion of the magnetic field with height, HMI underestimates the field strength and inclination in QS and plage regions but still captures the overall spatial distribution and orientation of magnetic structures \citep{2017_SainzDalda, 2025_Beck_Prasad_Hu_etal}.
In strong field regions, plasma-$\beta$ is low in the photosphere, so the field is magnetically dominant \citep{2021_Wiegelmann_Sakurai}, and HMI measurements show a better correlation.

Analysis of the filtered spatial coherence-weighted phase difference maps reveals that AGWs are efficiently suppressed in regions of intermediate to strong, vertically oriented fields, but propagate relatively unhindered in QS regions and predominantly transverse fields.
While there are phase suppression signatures in intermediate to strong fields in DS0 (Fig.\,\ref{fig:IBIS_spatial_distribution_phases}(a)), AGWs appear largely unaffected by the average QS magnetic field.
Positive phase signatures, indicative of propagating AGWs carrying energy downwards, and phase suppression become more pronounced with increasing magnetic structure size, like in sunspots (Fig.\,\ref{fig:IBIS_spatial_distribution_phases}(c1)--(c2)) and plage regions (Fig.\,\ref{fig:IBIS_spatial_distribution_phases}(d1)--(d2)), where stronger and more spatially extended magnetic fields alter and/or inhibit AGW propagation.
In particular, we note the striking resemblance of the propagation of AGWs in Fig.\,\ref{fig:IBIS_spatial_distribution_phases}(c1) in sensing the average magnetic field \added{structure,} matching the HMI LOS magnetogram for DS2 in Fig.\,\ref{fig:AGW2_reference_image}.

Spatial coherence-weighted phase difference maps for the filtered lower photospheric diagnostics primarily show minimal to no correlation with the magnetic field, indicating that AGWs are not substantially affected by the magnetic field \added{configuration} in the low photosphere, as suggested by \citet{2019_Vigeesh_Roth_Steiner_Jackiewicz}.
Additionally, analysis of the spatial power maps for the filtered diagnostics confirms power suppression and/or deficit in intermediate to strong, vertically oriented fields.

\subsubsection{Comparison with Sun-v100} \label{subsubsec:spatial_cobold}
\begin{figure*}
    \centering
    \includegraphics[width=\linewidth]{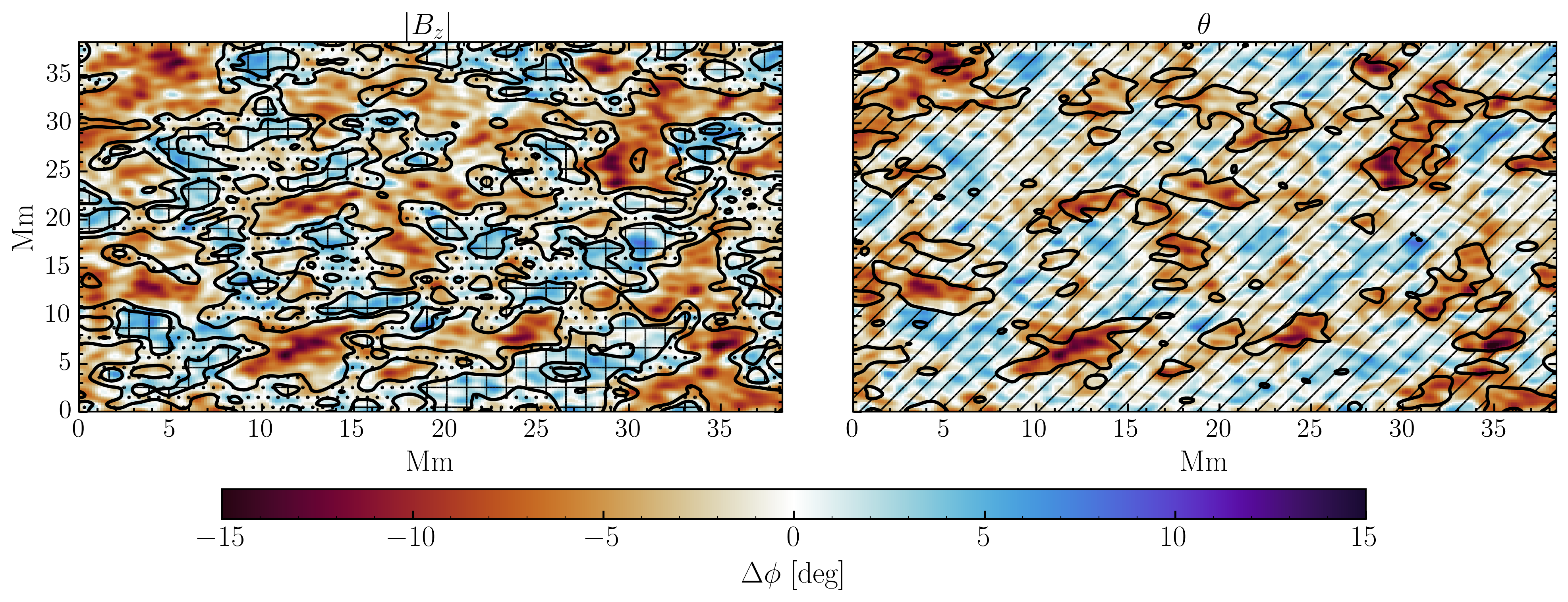}
    \caption{Spatial coherence-weighted phase difference maps for Sun-v100 at a height combination of 500--400\,km for AGWs, integrated within 0.7--3.1\,mHz, overlaid with the temporally averaged (left) unsigned magnetic field and (right) inclination contours from the lower height. These heights were chosen to replicate approximate average QS formation heights of \IBISFeUpperPhot{} and \IBISK{}. Contours are similar to those in Fig.\,\ref{fig:IBIS_spatial_distribution_phases}. Areas outside hatched contours correspond to QS and predominantly horizontal fields.}
    \label{fig:COBOLD_spatial_distribution_phases}
\end{figure*}

We compare our observational results with the Sun-v100 CO$^{5}$BOLD simulation, which features predominantly strong vertical fields.
We carry out a similar procedure, including filtering, to compute the spatial coherence-weighted phase difference maps for simulated heights of $400$\,km and $500$\,km to match the upper atmospheric IBIS diagnostics, which can be seen in Fig.\,\ref{fig:COBOLD_spatial_distribution_phases}.
The contours for the field properties are extracted from the field at $400$\,km.
Once again, regions outside the hatched areas represent QS fields and predominantly transverse fields.

The results agree remarkably well with our observational results in Fig.\,\ref{fig:IBIS_spatial_distribution_phases}.
In general, AGWs are efficiently suppressed or reflected in intermediate to strong, vertically oriented fields while they are allowed to propagate relatively unhindered in QS and transverse fields.
Additionally, although not shown, the magnetic field and inclination contours at the HMI sampling height revealed a somewhat less precise but still consistent correlation with phase, reinforcing our observational findings.

\subsubsection{Binned Coherence-weighted Phase Differences as a Function of Field Strength and Inclination} \label{subsubsec:binned_phases}

\begin{figure*}[hbt!]
    \centering
    \includegraphics[width=\linewidth]{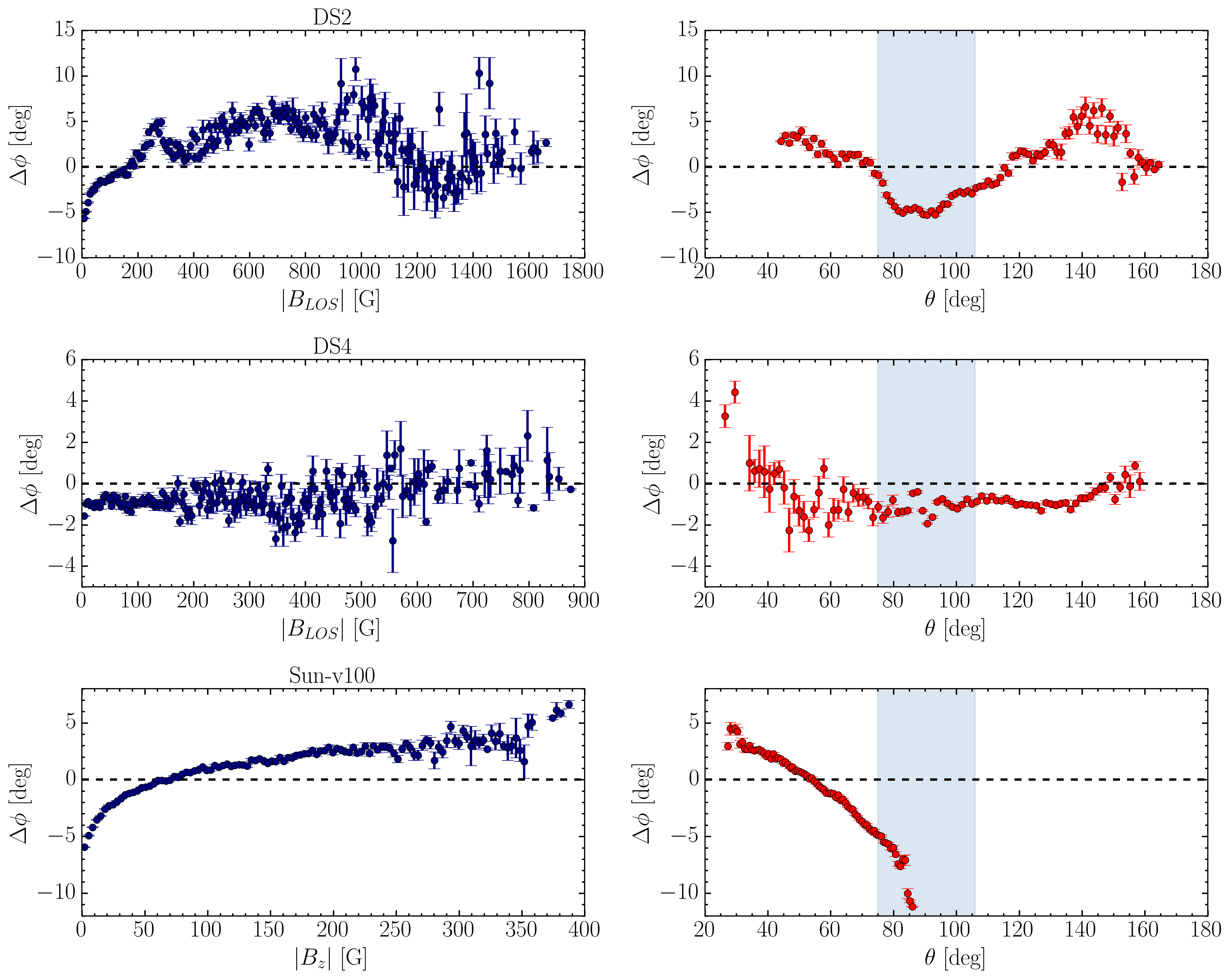}
    \caption{Binned coherence-weighted phase differences of the filtered AGWs with standard errors as a function of the (left panel) temporally averaged $|B_{LOS}|$ or $|B_z|$ and (right panel) field inclination ($\theta$) for DS2, DS4, and Sun-v100. The shaded blue regions on the right correspond to transverse fields ($75\degree$\,$\leq$\,$\theta$\,$\leq$\,$106\degree$).}
    \label{fig:binned_phases}
\end{figure*}

The effects of the magnetic \added{field configuration} on the propagation of AGWs can also be seen in the binned coherence-weighted phase differences as a function of the temporally averaged $|B_{LOS}|$ (left panel) and field inclination (right panel) for DS2, DS4, and Sun-v100, which are shown in Fig.\,\ref{fig:binned_phases}.
The dependence of the average field strength and inclination on the propagation of AGWs in the observations qualitatively matches that shown in Sun-v100, including the magnitude of the observed phase differences.

Phase differences become increasingly positive with stronger and vertically oriented fields, while they remain predominantly negative in QS and transverse fields.
This relationship is most evident for DS2, which matches well with the spatial distribution of phases seen in Fig.\,\ref{fig:IBIS_spatial_distribution_phases}(c1)--(c2).
In DS4, the trend is less distinct, likely due to field inhomogeneity present in plage regions (see Fig.\,\ref{fig:IBIS_spatial_distribution_phases}(d1)--(d2)).

Analysis of the field inclination as a function of field strength for DS4 shows that most negative phase differences are associated with $|B| < 160$\,G and a broad range of field azimuths.
In DS4, AGWs can propagate upwards to some extent for a range of field inclinations, including vertically oriented fields.
However, these phase differences are smaller in magnitude than DS2 or Sun-v100, suggesting complex wave interactions with the magnetic field.
We also note a difference in phase differences observed between outwardly and inwardly directed vertical magnetic fields.

\section{Discussion} \label{sec:discussion}

Our analysis reveals that AGWs are modulated by the magnetic field \added{geometry}, with these effects becoming increasingly apparent in regions with stronger, coherent, and organized magnetic fields, such as sunspots and plages.
The filtered spatial coherence-weighted phase differences with the magnetic field and inclination contours superimposed as well as the binned coherence-weighted phase differences as a function of field strength and inclination in Fig.\,\ref{fig:IBIS_spatial_distribution_phases} and Fig.\,\ref{fig:binned_phases}, respectively, reveal a trend of positive phase differences and phase suppression with increasing field strength and vertical inclination.
These results also qualitatively match the Sun-v100 CO$^{5}$BOLD simulation in Fig.\,\ref{fig:COBOLD_spatial_distribution_phases} and Fig.\,\ref{fig:binned_phases} remarkably well.

In general, the observed behavior is consistent with numerical simulations, where we corroborate that AGWs are suppressed and/or reflected in predominantly strong, vertical fields \citep[e.g.,][]{2017_Vigeesh, 2019_Vigeesh_Roth_Steiner_Jackiewicz} but propagate largely unhindered in QS and transverse fields \citep{2009_Newington_Cally, 2021_Vigeesh_Roth_Steiner_Fleck} in the upper photosphere.
Moreover, the striking resemblance between the AGW propagation patterns in Fig.\,\ref{fig:IBIS_spatial_distribution_phases}(c1) and the average magnetic field \added{structure} shown in the HMI magnetogram for DS2 in Fig.\,\ref{fig:AGW2_reference_image} highlights the magneto-seismology diagnostic potential of AGWs for probing the average magnetic field in the lower solar atmosphere.

The complexity of the interaction between AGWs and the magnetic field is evident through both observations and numerical studies.
Mode conversion and transmission are strongly influenced by the attack angle, which is the angle between the wave vector and the field direction \citep{2006_Schunker_Cally}.
As a result, even QS regions with highly inclined fields can cause wave reflection or conversion if the local geometry is favorable \citep[e.g.,][]{2011_Newington_Cally, 2016_Hague_Erdelyi}.
The simulated phase difference diagrams in Fig.\,\ref{fig:mag_diagnostic_diagrams} reinforce that the behavior of AGWs is highly affected by the geometry of the magnetic field.

In QS fields and areas with predominantly transverse magnetic fields, we observe significant negative phase differences, indicative of propagating AGWs carrying energy upwards.
Observational studies indicate that transverse fields are commonly observed in the QS \citep{2008_Lites_Kubo_SocasNavarro_etal}; therefore, this behavior is not unexpected.
The observed behavior agrees with numerical simulations showing that AGWs freely propagate upward in high plasma-$\beta$ environments, particularly transverse fields \citep[e.g.,][]{2017_Vigeesh, 2021_Vigeesh_Roth_Steiner_Fleck}.

This is also consistent with the framework by \citet{2009_Newington_Cally, 2011_Newington_Cally}, which proposes that when propagating AGWs are aligned with highly inclined fields (e.g., small attack angles), they can propagate up to the equipartition level relatively unscathed.
From there, they can transmit upward to higher atmospheric regions through a ``magnetic portal'' via mode conversion, similarly to the mechanism described by \citet{2006_Jefferies_McIntosh_etal}.
At this point, AGWs can undergo mode conversion into field-aligned acoustic or Alfvén waves \citep{2011_Newington_Cally_Proceeding}.
Partial mode conversion, transmission, and reflection are also probable, particularly in areas with steep temperature increases \citep{2012_Khomenko_Cally, 2012_Felipe}.
Temperature variations in the lower solar atmosphere could also create cavities where AGWs become trapped and turn to standing modes \citep{2016_Hague_Erdelyi}.
Therefore, suppressed and positive phase differences in Fig.\,\ref{fig:IBIS_spatial_distribution_phases} may be understood as signatures of reflected AGWs as slow modes from higher regions not sampled in our observations or the result of interference between upward and partially reflected propagating modes \citep{2009_Newington_Cally, 2012_Felipe}.

Subtle but systematic differences are also present between outwardly and inwardly directed fields of comparable inclination, as seen in Fig.\,\ref{fig:binned_phases}.
At inclinations greater than 106\textdegree, AGWs encounter increasingly unfavorable attack angles (the wave vector and field direction are antiparallel).
This leads to suppressed transmission and enhanced reflection or mode conversion, as noted in \citet{2009_Newington_Cally, 2024_Cally_Bogdan}.

In DS4, the negative phases become suppressed with increasing inclination and become positive beyond $\approx$\,150\textdegree, indicating a gradual transition from wave transmission to enhanced reflection.
This shift likely reflects the underlying magnetic \added{field structure} as observations show that the plage magnetic field is fragmented and spatially inhomogeneous, consisting of small-scale magnetic flux concentrations that expand with height into an inclined, canopy structure forming as low as 400--600\,km \citep{2015_Buehler_Lagg_Solanki_vanNoort, 2020_Morosin_delaCruzRodriguez_Vissers_Yadav}.
Observed opposite polarity patches beneath the canopy further underscore the complexity of the plage magnetic structure, which may contribute to partial reflection and phase variability \citep{2025_Liu_etal}.

In contrast, in DS2, the turning point that onsets phase suppression is more evident around 120\textdegree, where the phases increase steadily with inclination up to $\approx$\,140\textdegree.
At inclinations below 74\textdegree, the behavior is roughly mirrored for both datasets, but the magnitude of the phases in DS2 is smaller.
In Sun-v100, we predominantly have upright vertical fields, so the turning point of positive to negative phases is around 60\textdegree, which agrees qualitatively with our observations.
This suggests a critical efficiency threshold for AGW propagation in different field regimes, which differs from acoustic waves \citep{2006_Schunker_Cally, 2012_Khomenko_Cally}.

In all cases, increasingly horizontal fields and presumably smaller attack angles favor upward AGW propagation.
Conversely, suppressed or positive phase differences are indicative of reflected AGWs, observed as downward propagating slow modes modified by gravity \citep{2009_Newington_Cally, 2016_Hague_Erdelyi}.
Notably, while the AGW phase is generally reduced in regions of intermediate to strong fields, the threshold for this transition varies.
In Sun-v100, phase reversal happens around 60\,G, whereas in DS2 and DS4 it does not emerge until field strengths exceed 200\,G (with larger variability for DS4).

Surprisingly, the influence of the average magnetic field \added{configuration} is not apparent in our observed $k_h-\nu$ phase difference diagrams, as seen in Fig.\,\ref{fig:vv_phase}.
This contradicts the numerical simulations of \citet{2017_Vigeesh, 2019_Vigeesh_Roth_Steiner_Jackiewicz} and the observed results of \citet{2023_Kumar_Jumar_Rajaguru_Mathew_Bayanna}, who reported phase suppression, reflection, and/or scattering in different magnetic field configurations in $k_{\rm{h}}-\nu$ space.
However, we note that the azimuthal averaging analysis by \citeauthor{2023_Kumar_Jumar_Rajaguru_Mathew_Bayanna} was performed over selected regions of interest within the overall field of view of the SDO intensity diagnostics, potentially introducing noise and local variability.
In contrast, our results are consistent with synthetic observations by \citet{2020_Vigeesh_Roth}, who noted that the effects of the magnetic field on AGWs in $k_{\rm{h}}-\nu$ diagrams are likely not discernible as those in numerical simulations by \citet{2019_Vigeesh_Roth_Steiner_Jackiewicz}.
We speculate that this discrepancy in our observations is likely due to the spatial averaging across QS and ARs, where weak field areas dominate by area and dilute the effects of the magnetic field.

Additionally, we find that some magnetized phase differences simulated in Fig.\,\ref{fig:mag_diagnostic_diagrams} show qualitative similarities to the observed data in Fig.\,\ref{fig:vv_phase}.
In particular, the AGW phase difference distribution simulated for combinations of $a = 0.4c_s$ and $\theta \geq 50\degree$ in Fig.\,\ref{fig:mag_diagnostic_diagrams} bears a strong resemblance to the observed \IBISFeUpperPhot{}--\IBISK{} phase distribution in DS0, DS3, and DS4.
However, it remains unclear to what extent these correspondences only reflect modifications in the field geometry on the AGW regime, as suggested by Eq.~\ref{eq:3dmhd_dispersion}. 
Numerous numerical and simulation studies have revealed complex magnetic field and wave interactions \citep[e.g.,][]{2003_Bogdan_Carlsson_etal}, whose observational signatures in $k_{\rm{h}}-\nu$ space are poorly understood and might not be taken into account by dispersion relations.
It is probable that in observational cases that appear consistent with theory, the spectral diagnostics probe atmospheric regions where such complex wave and field interactions (e.g., reflection, mode conversion) are less prevalent. A similar explanation was proposed in \citet{1992_Deubner, 1996_Deubner_Waldschik}, where differing phase diagrams were attributed to variations in the sampled atmospheric regions, with some dominated by reflecting structures, others not. In these cases, it may be feasible to infer average field properties from the lower AGW boundary seen in the simulated $k_{\rm{h}}-\nu$ diagrams in Subsec.~\ref{subsubsec:magnetic_dispersions}, which will be explored in a future study.

We also sought to investigate the signatures of AGWs near the limb.
Given their predominately transverse nature, AGWs are expected to exhibit stronger horizontal (5--6\,km\,s$^{-1}$) than vertical (1--2\,km\,s$^{-1}$) velocities; therefore, their velocity signatures are predicted to be more prominent toward the limb \citep{1981_Mihalas_Toomre}.
\citet{1992_Deubner} reported enhanced negative phases in a QS dataset near $\mu=0.8$ between an upper photospheric and lower chromospheric diagnostic pair, which they interpreted as evidence of obliquely propagating AGWs.
However, we do not observe similar behavior, even in DS1 ($\mu = 0.57$) shown in Fig.\,\ref{fig:vv_phase}(a)--(b), although we do note larger velocity amplitudes in Fig.\,\ref{fig:AGW2_amplitude_spectral_densities}.
This likely reflects a combination of LOS projection effects, spectral smearing, and interference from incoherent wavefronts, which might obscure stronger phase signatures despite their strong amplitudes toward the limb.

\section{Conclusion} \label{sec:conclusion}

The detection of propagating AGWs for all spectral diagnostics, viewing geometries, and magnetic field configurations at the expected temporal and spatial scales in line with independent observations, numerical simulations, and theory indicates that AGWs are ubiquitous oscillations in the lower solar atmosphere.
Our results corroborate that AGWs are influenced by the average magnetic field \added{configuration}, particularly in the upper photosphere \citep{2009_Newington_Cally, 2017_Vigeesh}.
While the $k_{\rm{h}}-\nu$ phase difference diagrams reveal no discernible effects of the magnetic field in agreement with \citet{2020_Vigeesh_Roth}, the
filtered spatial coherence-weighted phase difference maps and the binned coherence-weighted phase differences reveal clear effects, such as suppression and reflection.
AGW propagation is highly affected by the magnetic field strength and inclination, where we observe that AGWs propagate relatively unhindered in QS and predominantly transverse fields.
In contrast, AGWs are efficiently suppressed or reflected in intermediate to strong vertical fields.

This work highlights the observational potential of using AGWs as magneto-seismology tools for probing the properties of the average magnetic field in the lower solar atmosphere.
The magnetic field can significantly alter wave propagation and energetics, with implications for local helioseismic inferences in magnetized regions \citep{2013_Cally_Moradi, 2016_Hague_Erdelyi}.
Therefore, multi-height diagnostics from future synoptic networks, such as the next-generation GONG network \citep[ngGONG;][]{2019_Hill_ngGONG}, especially in different field environments, will provide critical constraints on wave propagation and mode conversion.
Additionally, high-resolution, narrowband imaging from the Visible Tunable Filter \citep[VTF;][]{2012_Kentischer} recently installed on the Daniel K. Inouye Solar Telescope \citep[DKIST;][]{2020_Rimmele_DKIST, 2021_Rast_DKIST} will be essential for capturing the fine-scale coupling between AGWs and magnetic fields.

\begin{acknowledgments}
The IBIS datasets used in this publication were obtained with the Dunn Solar Telescope facility, operated by New Mexico State University with funding support from the National Science Foundation and the state of New Mexico. We thank the DST staff for their help in taking these observations, particularly Doug Gilliam. The HMI data used in this publication are courtesy of NASA's SDO and the HMI science team. O.V. acknowledges that this work was supported by the COFFIES DSC Cooperative Agreement 80NSSC22M0162. J.M. is supported by a subcontract of the COFFIES DSC to NMSU.
\end{acknowledgments}

\facilities{Dunn(IBIS), SDO(HMI)}

\software{CMasher \citep{2020_JOSS_cmasher}}

\bibliography{agw2}{}
\bibliographystyle{aasjournal}

\end{document}